\documentclass{aa}
\pdfoutput=1
\usepackage{graphicx}
\usepackage{epsfig}
\usepackage{txfonts}
\usepackage{wasysym}
\usepackage{marvosym}

\def\mearth{M_\oplus}
\def\msun{M_\odot}

\def\mcore{M_{\rm core}}
\def\mheavy{M_{\rm heavy}}
\def\mztot{M_{\rm Z,tot}}
\def\astart{a_{\rm start}}
\def\fpg{f_{\rm D/G}} 

\def\f1{f_{\rm I}}
\def\mj{M_{\textrm{\jupiter }}}
\def\mstar{M_*}

\def\astart{a_{\rm start}}
\def\tstart{t_{\rm start}}
\def\membstart{M_{\rm emb,0}}

\def\anorm{a_0}

\def\mwind{\dot{M}_{\rm w}}
\def\menv{M_{\rm env}}
\def\miso{M_{\rm iso}}

\def\arock{a_{\rm rock}}

\def\beq{\begin{equation}}
\def\eeq{\end{equation}}

\def\atouch{a_{\rm touch}}
\def\sigmanorm{\Sigma_{0}}
\def\tmc{\tau_{\rm RV,MC}}
\def\epsilonmc{\epsilon_{\rm RV,MC}}
\def\nsynt{N_{\rm synt}}
\def\ninit{N_{\rm init}}
\def\nnocalc{N_{\rm nocalc}}
\def\nobssynt{N_{\rm obssynt}}
\def\nhot{N_{\rm hot}}
\def\nobsreal{N_{\rm obsreal}}
\def\nsamp{N_{\rm samp}}
\def\dks{d_{\rm KS}}
\def\sam{S_{\rm a-M}}
\def\sm{S_{\rm M}}
\def\sa{S_{\rm a}}
\def\sfeh{S_{\rm [Fe/H]}}

\def\msini{M\sin i}

\def\pwhot{P_{\rm whot}}
\def\pfv{P_{\rm FV}}
\def\pfvwhot{P_{\rm FV,whot}}
\def\ffv{F_{\rm FV}}

\def\mmax{M_{\rm max}}

\def\aj{AJ}                   
\def\araa{ARA\&A}             
\def\apj{ApJ}                 
\def\apjl{ApJ}                
\def\apjs{ApJS}               
\def\ao{Appl.Optics}          
\def\aap{A\&A}                
\def\mnras{MNRAS}             
\def\pasp{PASP}               

\def\({\left(}
\def\){\right)}
\def\<{\left<}
\def\>{\right>}

\begin{document}

\title{Extrasolar planet population synthesis II:\\ Statistical comparison with observation}

\author{Christoph Mordasini\inst{1,2}  \and Yann Alibert\inst{1,3}  \and Willy Benz \inst{1} \and Dominique Naef\inst{4}}

\institute{Physikalisches Institut, University of Bern, Sidlerstrasse 5, CH-3012 Bern, Switzerland \and  
Current address: Max-Planck-Institut f\"ur Astronomie, K\"onigstuhl 17, D-69117 Heidelberg, Germany \and
Institut UTINAM, CNRS-UMR 6213, Observatoire de Besan\c{c}on, BP 1615, 25010 Besan\c{c}on Cedex, France \and
 European Southern Observatory, Alonso de Cordova 3107, Casilla 19001 Santiago 19, Chile} 

\offprints{Christoph MORDASINI, \email{mordasini@mpia.de}}

\date{Received 28 July 2008 / Accepted 15 April 2009}

\abstract
{This is the second paper in a series of papers showing the results of extrasolar planet population synthesis calculations using our extended core accretion model. In the companion paper (Paper I), we have presented in detail the methods we use. In subsequent papers, we shall discuss the effect of the host star's mass on the planetary population and the influence of various properties of protoplanetary disks.}
{In this second paper, we focus  on planets orbiting solar-like stars.  The goal is to use the main characteristics of the actually observed extrasolar planet population to derive in a statistical manner constraints on the planet formation models.}
{Drawing initial conditions for our models at random from probability distributions derived as closely as possible from observations, we synthesize  a number of planetary populations. By applying an observational detection bias appropriate for radial velocity surveys, we identify the potentially detectable synthetic planets. The properties of these planets are compared in quantitative statistical tests with the properties of a carefully selected sub-population of actually observed extrasolar planets.}
{We use a two dimensional Kolmogorov-Smirnov test to compare the mass-distance distributions of synthetic and observed planets, as well as the one dimensional version of the test to compare the $\msini$, the semimajor axis and the [Fe/H] distribution. We find that while many combinations of parameters lead to unacceptable distributions, a number of models can account to a reasonable degree of statistical significance for most of the properties of the observed sample. We concurrently account for  many other observed features, e.g. the ``metallicity effect''. This  gives us confidence that our model captures several essential features of giant planet formation. In addition, the fact that many parameter combination could be rejected, indicates that planet population synthesis is indeed a promising approach to constrain formation models. Our simulations allow us also to extract a number of properties of the underlying exoplanet population that are not yet directly detectable. For example, we have derived the planetary initial mass function (PIMF) and have been led to conclude that the planets detected so far represent only the tip of the iceberg (9\%) of all the existing planets. The PIMF can also be used to predict  how the detectable extrasolar planet population will change as the instrumental precision of radial velocity surveys improves from $\sim$ 10 m/s to $\sim$1 m/s, or even to an extreme precision of 0.1 m/s.}{}


\titlerunning{Extrasolar planet population synthesis II}
\authorrunning{C. Mordasini et al.}

\maketitle
\section{Introduction}\label{sect:introduction}
In the first paper of this series (Mordasini et al. \cite{paperI}, hereafter Paper I), we have presented our methods to synthesize populations of extrasolar planets. We have explained how we use our extended core accretion model (Alibert et al. \cite{alibertetal2005a}) to generate synthetic planetary populations by varying in a Monte Carlo fashion four key variables describing the initial conditions in our planet formation model. As shown in Paper I, we have tried in deriving the probability distribution of these four variables to stay as close as possible to actual observations. 

We have found that the large spread of initial conditions resulting from the variation of the characteristics of the proto-planetary disk (abundance of heavy elements, mass and lifetime) and their relative probability of occurrence leads to the formation of a synthetic population of planets characterized by a large diversity. Hence, we argued that, within the core accretion paradigm, the observed diversity of exoplanets is a natural consequence of the diversity of disk properties. 

In Paper I, we have also identified a number of typical phases planets undergo during their formation, and found that these phases lead to characteristic planetary formation tracks. These tracks determine the final position of each planet in the distance to star versus planetary mass diagram (a-M) and therefore can be used in order to interpret the corresponding observational diagram. 

Unfortunately, not all model parameters can be constrained by observations of proto-stellar disks. To circumvent this problem, we present in this paper an approach that consists in comparing statistically the overall characteristics of our synthetic planets with those of a carefully selected sub-population of actually detected exoplanets. This approach has been made possible by the large number of exoplanets that have been detected over the recent years which has allowed to go beyond the characteristics of individual objects and define the characteristics of the ensemble population. Many studies have discussed from a observational point of view the statistical properties of the extrasolar planets, analyzing various distributions and correlations in order to address the following (and many more) issues, as recently reviewed by Udry \& Santos (\cite{udrysantos2007}).

(1) Before the detection of 51 Peg b 13 years ago by Mayor \& Queloz (\cite{mayorqueloz1995}) it was not clear if planets outside our own Solar System existed, although from a theoretical point of view, there was no reason to doubt it. Nowadays we know that roughly 5-10 \% (e.g. Marcy et al. \cite{marcyetal2005}; Cumming et al. \cite{cummingetal2008}) of solar-like star in the solar neighborhood harbor a giant planet within a few AU in distance.

(2) Detection biases still hinder the exploration of the full planetary mass domain. It is however clear that the mass distribution increases towards small mass planets (e.g. Butler et al. \cite{butleretal2006}; Jorissen et al. \cite{jorissenetal2001}), which points towards the existence of a large number of yet undetected low mass planets. It is also known that there are very few objects with masses larger than $\sim15$ Jupiter masses inside a few AU (e.g. Marcy \& Butler \cite{marcybutler2000}) defining the ``brown dwarf desert''. With the detection of smaller and smaller mass planets, new, finer structures in the mass distribution, like a bimodal shape at very low masses (Mayor et al. \cite{mayoretal2008}) have recently been suggested.

(3) The distribution of semimajor axes consists of a pile up of Hot Jupiters at about 0.03 AU, followed by a relative depletion (the ``period valley'') and finally an increase in frequency further out at  about 1 AU (e.g. Udry et al. \cite{udryetal2003}). Outside a few AU the limited time duration of the surveys does not allow definitive statements yet. 

(4) The combination of mass and distance has shown that there is an absence of massive planets at small orbital distances (e.g. Zucker \& Mazeh \cite{zuckermazeh2002}), and a positive correlation of planetary mass and distance (e.g. Jiang et al. \cite{jiangetal}). Low mass, Neptunian planets seem to be characterized by a different distribution than giant planets (Udry \& Santos \cite{udrysantos2007}).

(5) Soon after the first discoveries of extrasolar planets it was noticed that the detection probability of giant planets increases with stellar metallicity (Gonzalez \cite{gonzalez1997}). This ``metallicity effect'' is now very well established (e.g. Fischer \& Valenti \cite{fischervalenti2005}; Santos et al. \cite{santosetal2003}). Also correlations between stellar metallicity and the planetary semimajor axis have been discussed (e.g. Sozzetti \cite{sozzetti2004}), but no definitive conclusions can be drawn at this time. The stellar mass certainly also plays a role for planet formation. Observations of stellar types other than FGK underly certain complications, but a positive correlation between stellar mass and frequency of massive planets seems now to be clear (e.g. Lovis \& Mayor \cite{lovismayor2007}).
 
(6) It was found that planets in relatively tight binary systems have statistically different properties (e.g. Eggenberger et al. \cite{eggenbergeretal2004}; Desidera \& Barbieri \cite{desiderabarbieri2007}). For example, stars in binaries have close-in very massive planets, absent around single stars (e.g. Zucker \& Mazeh \cite{zuckermazeh2002}).  This points toward a possible role of the environment in planet formation. 

(7) The observed population is now known to have an eccentricity distribution that is similar to the one of stellar binaries, although a group of long period, low eccentricity giant extrasolar planets more similar to our giants exists (e.g. Halbwachs et al. \cite{halbwachsetal2005}). 
 
In this work, we are addressing the points 1 to 5 while the remaining ones are for the moment beyond the capabilities of our model. In particular we wanted to investigate if by varying the otherwise unconstrained model parameters over a reasonable range of values, it was possible to reproduce, in a statistical significant manner, as many as possible of the characteristics of the observed exoplanet population listed above. To do so requires the ability to \textit{``observe''} the synthetic planets with the same detection biases as actual observations in order to extract the directly comparable sub-population. Since we use as the observational comparison sample planets that have been detected by radial velocity (RV) methods, we have used the detection biases relevant for this type of planet searches with different intrinsic instrumental precisions. A similar approach can be used for any type of detection biases relevant to different detection techniques (transits, lensing, astrometry, direct imaging). Since these other techniques are sensitive to other planet characteristics than RV techniques, such comparisons would provide additional and independent checks of the model which we intend to carry out in future work. A first such example is shown in \S \ref{subsubsect:heavyelementhots} where we compare the amount of heavy elements of close-in synthetic planets with the one derived from internal structure modeling of transiting Hot Jupiters (Guillot \cite{guillot2008}).  

Another distinct advantage of population synthesis calculations is that they allow to study the global consequences  of a certain physical mechanism as shown by Ida \& Lin (\cite{idalin2008b}) for the example of ``dead zones''. In this paper, as we concentrate on the comparison with the observed population, we only discuss two such effects, namely the absence of solids inside a given semimajor axis and type I migration, and postpone further studies to subsequent publications.   

The outline of the paper is as follows: In section \S\ref{sect:methodsshort} we summarize very shortly the methods described in Paper I we use to obtain the planetary populations.  Section \S\ref{sect:detectionbiases} describes the procedure to sort out  the detectable synthetic planets from the whole population using a synthetic observational bias. The following section \S\ref{sect:statisticalanalysis} shows how we have statistically compared this sub-population with the real  exoplanets. The results concerning this comparison are given in section \S\ref{sect:compresults}, while those concerning the predictions for extremely precise radial velocity survey are in \S\ref{sect:predictionsRV}. The conclusions are drawn in the last section, \S\ref{sect:conclusions}.      

\section{Methods}\label{sect:methodsshort}
In Paper I, we have described in detail the six step method we use to synthesize extrasolar planet populations. We have in particular described the (small) changes we made to our extended core accretion model that was presented in Alibert et al. (\cite{alibertetal2005a}), necessary to allow for the very large number of calculations occurring for population synthesis. In our model we solve as in classical core accretion models (e.g. Pollack et al. \cite{pollacketal1996}) the internal structure equations for the forming giant planet, but at the same time we include disk evolution (using the $\alpha$ formalism, see Papaloizou \& Terquem \cite{PT99}) and type I and II planetary migration.

We have then described the four Monte Carlo variables that describe the varying initial conditions for planet formation, explaining in particular how we have derived their probability distributions from observations of (mainly) circumstellar disks. The random variables are (1) the dust-to-gas ratio $\fpg$ which is constrained by observed stellar metallicities [Fe/H] (Murray et al. \cite{murray2001}; Santos et al. \cite{santosetal2003}), (2) the initial gas surface density  $\sigmanorm$ at $\anorm=5.2$ AU which is constrained by observed disk masses (Beckwith \& Sargent \cite{beckwith1996}), (3) the rate at which photoevaporation occurs $\mwind$, which determines together with $\alpha$ the disk lifetime, and is therefore constrained by observations of Haisch et al. (\cite{haischetal2001}), and (4) the starting position of the embryo $\astart$. Planetary seeds are allowed to start only in these parts of the disk where the isolation mass $\miso$ is larger than the initial embryo mass $\membstart=0.6 \mearth$ and where the starting time $\tstart$ of an embryo which is the time needed to build up such an objects is shorter than the disk lifetime.

Each population is also characterized by a number of parameters that are kept constant for all planets. The most important (and worst constrained) parameters of the model are the viscosity parameter $\alpha$ for the gas disk and the efficiency factor for type I migration $\f1$. By synthesizing populations using various combinations of $\alpha$ and $\f1$ (\S\ref{subsect:influeceofparameters}), and comparing them to the observed population,  we have found that  the model with $\alpha = 7\times10^{-3}$ and $\f1 = 0.001$ provides the overall best statistical results. We define this model as the nominal model. This population was already presented and discussed in Paper I.

For each planet we record the full time evolution from the seed embryo to its final mass and position (cf. planetary formation tracks in Paper I). For the statistical analysis presented here,  we only use the final characteristics of  the planets:  The final semimajor axis $a$, the total mass of accreted planetesimals $\mheavy$, the mass of the envelope $\menv$, the total mass $M$, and the formation time of the planet. Note however that we also posses the fractions of icy and rocky material that were accreted during the formation, which allows us to study certain objects as \object{GJ 436 b} in detail  (Figueira et al. \cite{figueiraetal2008}). To be able to compare quantitatively with radial velocity observations we also compute projected masses $\msini$ for which we assume a random orientation of extrasolar planetary orbits relative to the Earth. 

\section{Detection biases}\label{sect:detectionbiases}
To statistically compare the characteristics of the synthetic population to the actually observed one, we need to go one step further. We must identify which subgroup of synthetic planets actually could be detected in an observational survey. For this purpose, it is necessary to  understand and quantify the various detection biases entering into the observational process and apply them to the synthetic set. Such biases are technique, instrument and very likely also observer dependent.  In this work, we have considered only the biases affecting the radial velocity (RV) technique, as so far the large majority of exoplanets has been detected using this technique. However, our approach can be applied to any observational technique for which a detection probability can be calculated as function of semimajor axis and mass or planetary radius.

\subsection{Synthetic RV bias}\label{subsection:syntheticRVbias}
To first order, the planet detection probability based on RV measurements increases with increasing planetary mass and decreasing distance. The instrumental precision $\epsilon_{RV}$ then determines whether the planet can be detected or not.  But in fact, a large number of other quantities also affect the detection probability: The magnitude of the star, its rotation rate, the orbital eccentricity of the planet, the actual measurement schedule, stellar jitter and more. Using the method originally developed by Naef et al. (\cite{naef2004}, \cite{naefetal2005}) for the spectrograph ELODIE,  we determine by a $\chi^2$ analysis in a two dimensional grid in planetary period  ($1\leq P \leq 40\,000$ days) and mass  ($1\leq M \leq 12\,720$ $\mearth$) on each, out of a total of $5612$ grid points, the fraction of $50\,000$ randomly chosen planetary orbits that can actually be detected by a spectrograph of a given precision $\epsilon_{RV}$, taking into account all the effects mentioned above (see Naef et al. \cite{naefetalinprep} for a detailed description).  This fraction represents the detection probability corresponding to a given planetary period and mass.

\begin{figure}
   \resizebox{\hsize}{!}{\includegraphics{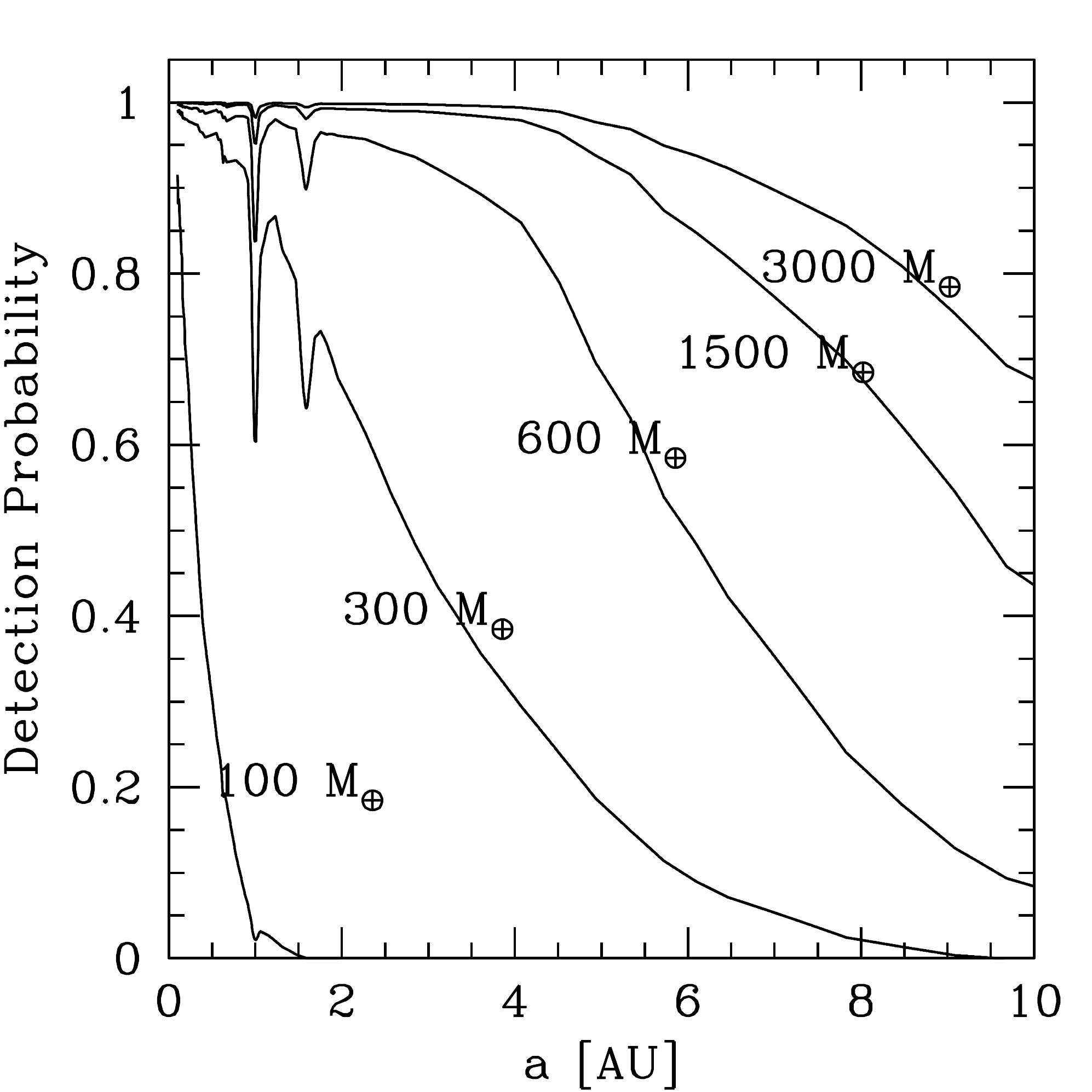}}
   \caption{The detection probability as a function of semimajor axis $a$ for five different planetary masses between 100 and 3000 $\mearth$ and an  instrumental accuracy of $\epsilon_{RV}$=10 m/s, as used for our synthetic RV survey. The stellar mass is 1 $\msun$, and it is assumed that the observations  completely cover at least one orbital period.} 
   \label{fig:detecprobRV}
\end{figure}

The results of these calculations for $\epsilon_{RV}$=10 m/s are illustrated in fig. \ref{fig:detecprobRV}. The graph shows the detection probability as a function of semimajor axis for a choice of five  planetary masses between 100 $\mearth$ and 3000 $\mearth$.  We see for example that planets with a mass higher than about five times  the mass of Jupiter $\mj$ ($\sim$ 1500 $\mearth$) can be detected with very high probability out to a distance of about five AU. For a planet of about one $\mj$, the detection probability falls below 50\% outside roughly 2.5 AU. Note also the stroboscopic effects at orbital periods of one and two years, leading to a reduction of the detection probability at the corresponding semimajor axes.
 
To decide if a synthetic planet would have been detected, we first determine its detection probability by interpolation in the detection probability grid, and  then draw a random number between 0 and 1. If the latter is smaller the detection probability, the planet counts as one of the $\nobssynt$ detectable synthetic planet and will be used for the statistical  tests (\S\ref{sect:statisticalanalysis}). 

\subsection{Synthetic RV survey}\label{subsect:syntheticRVsurvey}
With the synthetic RV detection bias at hand, we can construct a synthetic RV survey by ``observing''  the planet population coming out of the model. To do so, we have to specify two quantities  that characterize our MC survey:  First its instrumental precision $\epsilonmc$, and second its temporal duration $\tmc$.  As we want to statistically compare the subgroup of detectable synthetic planets with the  real observations, these two quantities should represent a  time and instrument average over the real RV surveys conducted by various  teams using various instruments over the last several years. Defining such averages is not trivial,  as the instrumental accuracy has changed from typically 10 m/s at the times of the discovery of 51 Peg b (Mayor \& Queloz \cite{mayorqueloz1995}) down to less than $1$ m/s with HARPS (Pepe et al. \cite{pepeetal2004}). This would have the effect that at smaller semimajor axes, planets of a smaller mass are known, so that $\epsilonmc$ also should be a function of time, and weighted by the contributions of the different observer teams.  For simplicity we have assumed for our synthetic MC survey a constant $\epsilonmc$=10 m/s and a survey duration of $\tmc$=10 years. The later is needed since generally planet discoveries are only announced when one full orbital period has been covered with observations (Cumming et al. \cite{cummingetal2008}). Therefore, our subgroup of potentially detectable synthetic planets contains only planets with an orbital period smaller than $\tmc=$10 yrs corresponding to a semimajor axis of about 4.6 AU.

In section \S\ref{sect:predictionsRV} we have studied the effects of  changing $\epsilonmc$ and $\tmc$, for which we have recalculated our bias tables for $\epsilonmc$=1 m/s and 0.1 m/s.

\subsection{Observational comparison sample} \label{subsection:observationalcomparisonsample}
Of the more than 300 currently known extrasolar planets, not all can be used for quantitative comparisons with our model, as some planets conflict with the fundamental assumptions on which the model is based (Alibert et al. \cite{alibertetal2005a}; Paper I). We therefore have to isolate the ones appropriate for the statistical test. In particular, we select extrasolar planets based on the following criteria:

\subsubsection{Sample selection criteria}
(1) The evolution of a planet may be significantly altered by the presence of another (massive) planet (Alibert et al. \cite{alibertetal2005b}, \cite{alibertnewsyst2006}; Thommes et al. \cite{thommesetal2008}). In contrast, we follow the evolution of just one embryo per disk (Paper I),  and therefore do not use any exoplanet that is member of a known extrasolar planetary system. It is clear that many single extrasolar planets could in fact be members of multiple systems with the small planets not detectable today. In this case, we argue that these small planets did not affect significantly the evolution of the massive one known today. 

(2) As explained in Paper I, our model doesn't describe adequately planets that migrate closer to the star than $\atouch\approx0.1$ AU.  For qualitative comparisons, we have thus to exclude all observed planets that fall into this $a-M$ domain. This especially means that Hot Jupiters are excluded, reducing significantly the number of comparison planets. The rate of occurrence of Hot Jupiters is however a too important constraint on migration to ignore it, so we still take it into account separately, as described in \S\ref{subsect:obsconsthotjup}.

(3) Planets in binary or multiple stellar systems have different statistical properties  than planets orbiting single stars (Eggenberger et al. \cite{eggenbergeretal2004}; Desidera \& Barbieri \cite{desiderabarbieri2007}). This could be due to different migration mechanism (Eggenberger et al. \cite{eggenbergeretal2004}).  We therefore do not consider planets in stellar systems with a binary separation less than 300 AU. Binaries with a wider separation than $\sim300$ AU have planets which exhibit no significant statistical differences from those around single stars  (Desidera \& Barbieri \cite{desiderabarbieri2007}).  

(4) In all simulations presented here, the stellar mass $\mstar$ is fixed to 1 $\msun$. For stars with masses not too different from the solar mass which are the primary targets of RV surveys, say for FGK stars which have masses between $0.7\lesssim\mstar/\msun\lesssim1.3$, no evident correlation between $\mstar$ and planetary properties have been found up to date (Udry \& Santos \cite{udrysantos2007}).  For example,  Fischer \& Valenti (\cite{fischervalenti2005}),  hereafter FV05 don't see a correlation between stellar mass and formation probability of gas giant planets for this small domain of primary masses. However, for stars with significantly different masses (M dwarfs, namely), theoretical studies (Laughlin et al. \cite{laughlinetal2004}; Ida \& Lin \cite{idalin2005}) indicate planetary populations with significantly different properties, containing for example less giant planets. Indeed, observations also indicate that giant planets are less frequent around M dwarfs. Ongoing dedicated M dwarf surveys e.g. with HARPS (Bonfils et al. \cite{bonfilsetal2005}) will help to constrain this issue further from an observational point of view. On the other hand,  intermediate mass stars ($1.5 \lesssim \mstar / \msun \lesssim 4$) seem to have more and more massive giant planets (Lovis \& Mayor \cite{lovismayor2007}).  Due to this reasoning, we only include planets of stars with $0.7<\mstar/\msun<1.3$ in our comparison.

(5) Many extrasolar planets have high, sometimes even very high eccentricities. In our model in contrast, synthetic planets can only have circular orbits (Paper I). Even though planet-disk interactions can pump eccentricity under certain circumstances too (Goldreich \& Sari \cite{goldreichsari2003}), the high eccentricities have been interpreted as mainly the result of gravitational interactions between (proto-)planets in initially more crowded systems (e.g. Rasio \& Ford \cite{rasioford1996}; Adams \& Laughlin \cite{adamslaughlin2003}; Veras \& Armitage \cite{verasarmitage2006}).  Thus, even if there is only one highly eccentric planet left today orbiting a certain star, its high eccentricity could be an indication that  the number of planets might have been larger during the formation epoch, and that planet-planet interaction have been important in the system. Hence, planets with a high eccentricity are more likely than low eccentricity planets to have conflicted during their formation with our criterion (1). We have therefore excluded planets with an eccentricity larger than $e_{max}=0.3$

(6) In the last few years, the accuracy of RV measurements has improved significantly, bringing the detection limit down to less than 1 m/s in some cases (Mayor et al. \cite{mayoretal2008}). Our synthetic RV survey has in contrast a precision $\epsilonmc=10$ m/s. We therefore use the same synthetic bias as for the synthetic planets (\S \ref{subsection:syntheticRVbias}) to sort out also these known exoplanets that could have been detected by our synthetic survey \textit{i.e.} we only consider real exoplanets with a period of less than 10 years and a sufficiently large mass to be detectable by our synthetic bias. 

Additionally, we only include planets detected by the radial velocity method, as other methods have a different detection bias. For the comparison sample of actual extrasolar planets, we use the compilation of observational data available online at J. Schneider's Extrasolar Planet Encyclopedia\footnote{\texttt{http://exoplanet.eu}}.

After applying the six criteria mentioned above, there are only $\nobsreal=$ 32 known extrasolar planets left. A larger number would obviously be very desirable, which shows the importance of persistent unbiased observational campaigns.  

\subsubsection{Representativity of the sample}
The low number also makes one wonder whether or not this small sub-sample represents well the overall sample of all known exoplanets. Our selection criteria in particular exclude planets which probably underwent strong planet-planet interactions (criteria 1 and 5).  Thommes et al. (\cite{thommesetal2008}) have shown that disks which lead to the formation of just one giant gaseous planet might only represent a very specific class of all planet-forming disks, namely those near the threshold for giant planet formation. When the mass of the disk is in contrast e.g. well above the threshold, prolific giant planet formation could occur. This could in turn be reflected in the final properties of the planets. It is therefore also from this perspective interesting to compare our 32 planet sample with a larger set of exoplanets.

To do so, we disregarded criterion 1 or 5, or both together  and studied if the associated $a-M$, $M$, $a$ and [Fe/H] distributions are significantly different from the $\nobsreal=32$ case. We find that this is in general not the case. We also find that for the differences that do occur, the eccentricity criterion is more important than the multiplicity criterion alone.  The only difference that is probably significant occurs for the mass distribution, with the more eccentric planets being more massive. Such a correlation among the known extrasolar planets was already pointed out some time ago by Marcy et al.  (\cite{marcyetal2005}), and can indeed be explained by planet-planet interactions: More massive disks produce more massive planets, and in higher numbers, leading to stronger scattering and therefore higher eccentricities, a behavior that is indeed  seen in the simulations of Thommes et al. (\cite{thommesetal2008}). This effect is even strengthened by the fact that in planet-planet interactions, the less massive bodies tend to get ejected. The visual impression that the two distributions are however not extremely different is confirmed by a Kolmogorov-Smirnov test (cf. the next section below) indicating a significance of still 37 \% that the two samples are identical. Considering the semimajor axis distribution, our 32 planet sample seems to contain slightly more planets at smaller distances ($a\lesssim 1 $ AU). This difference is however clearly not statistically significant. For the metallicity, no clear trends are visible either. One can see an absence of very high eccentricity ($\gtrsim0.6$) planets at low [Fe/H]$\lesssim-0.1$. But the KS significances are always higher than 85 \% (as for $a$) that the samples are identical, so this correlation is with the still low number of planets not significant. The general lack of significant correlations between $a$, $e$ and [Fe/H] has already been found elsewhere (FV05; Udry \& Santos \cite{udrysantos2007}).

We conclude that in general our comparison sample, despite being small, represents quite well the overall giant planet population around solar like stars, except for a likely shift to somewhat smaller masses. This result could indicate that at least for giant planets, our one-embryo-per-disk approach (see Paper I) leads in a statistical sense not to completely different results than the real multi-body formation process (Thommes et al. \cite{thommesetal2008}). For smaller mass planets, such a generalization might however be less well-founded.

\section{Statistical analysis}\label{sect:statisticalanalysis}
To assess the statistical significance of our results we perform four Kolmogorov-Smirnov (KS) tests in which we compare the distributions of the most important properties of the potentially detectable synthetic and the real planets. The null hypothesis is that the synthetic potentially detectable planets and the real planets are drawn from the same parent distribution. Small values of the significance level $S$ returned by the test show that this null hypothesis has to be rejected with a high probability of $1-S$. In particular, we perform three 1-dimensional tests (for the distributions of $\msini$, $a$ and [Fe/H]), and one 2-dimensional KS test in the $a-\msini$ plane.  

In the 1-D case, we run standard two sided KS tests (Press et al. \cite{pressetal1992}), comparing the sub-population of $\nobssynt$ detectable synthetic planets  with the observational comparison sample containing $\nobsreal=32$ real extrasolar planets. In 1-D it is possible to directly calculate the significance level $S$ once the KS distance $\dks$ is known (Press et al. \cite{pressetal1992}) as 
\begin{eqnarray}\label{eq:significanceKSfromdKS1}
N_{e}&=&\frac{\nobsreal \nobssynt}{\nobsreal+\nobssynt} \\
x&=&(\sqrt{N_{e}}+0.12+0.11/\sqrt{N_{e}})\times \dks\\
S(x)& = &2 \sum_{j=1}^\infty (-1)^{j-1} \exp(-2 j^2 x^2). \label{eq:significanceKSfromdKS3}
\end{eqnarray}	

In 2-D,  the distribution of the KS distances $\dks$ for the null hypothesis is not independent of the shape of the distributions that are being examined (Press et al. \cite{pressetal1992}), and thus the analytical transformation of the $\dks$ values to the significance $S$ is not accurate in all cases. Therefore, we have proceeded as described by Press et al. (\cite{pressetal1992}) in order to get the significance directly. After having generated a large number of detectable synthetic planets  ($\nobssynt$), that serve as the synthetic comparison population, we generate another $\nsamp$ samples of synthetic detectable planets, each one containing the same number of synthetic planets $\nobsreal=32$ as the real observational comparison sample. For each of these synthetic bootstrap samples we compute their KS distance $\dks$ to the $\nobssynt$ planets of the detectable synthetic sub-population. Finally we also compute the KS distance of the real observations, and calculate the fraction of cases of $\nsamp$ where these synthetic $\dks$ exceed the $\dks$ from the real data. This fraction is then the significance $S$.  

For each bootstrap sample we also compute as a check the three KS distances $\dks$ for the 1-D tests with this procedure, so that we also get the three significances for $\msini$, $a$ and [Fe/H] in this direct way without using the equations above that link $\dks$ and $S$. As expected for the 1-D case, the two methods always yield very similar results. 

\subsection{Defining observational constraints on the results}\label{subsect:definingobsconstarints}
One of our main goals is to test if it is possible to reproduce \textit{all} (or at least the most important) observational characteristics \textit{at the same time} with \textit{one} single synthetic population. We considered the following six observational characteristics as constraints to the model: (1) A high statistical KS significance $\sam$ for the two dimensional distribution in the $a-\msini$ plane (\S\ref{subsect:aMdiagram}), idem for the one dimensional distributions of  (2) the mass $\sm$ (\S\ref{subsect:msini}), (3) the semimajor axis $\sa$ (\S\ref{subsect:semimajoraxis}), and (4) the metallicity $\sfeh$ (\S\ref{subsect:metallicity}), then (5) a Hot Jupiter fraction $\ffv$ which is compatible with observation (\S\ref{subsect:obsconsthotjup}), and finally (6) a correct reproduction of the ``metallicity effect'', \textit{i.e.} the increase of the detection probability with stellar metallicity (\S\ref{subsect:metallicityeffect}).    

We also compared the overall detection probability $P$ of our synthetic survey (fraction of embryos that grew to become detectable planets) with the actual values, but we should bear in mind that our Monte Carlo simulations yield strictly speaking a different result than the observations due to the one-embryo-per-disk simplification: We can calculate the probability that one specific embryo with a given $\astart$ and $\tstart$ becomes detectable, whereas observations yield the fraction of stars for which any of the initially numerous embryos in the disk finally became a detectable (giant) planet.  

To see if we can fulfill the observational constraints, we have generated several populations, keeping the probability distributions which are constrained by observations fixed, but varying some parameters (mainly $\alpha$ and $\f1$). The varied parameters are listed in table \ref{tab:variedparameters}, and their influence is discussed in \S\ref{subsect:influeceofparameters}.  Typically, we were confronted with the fact that changing one parameter had multiple effects, bringing our results closer to one observational constraint, while at the same time the results for another deteriorated. However, many combinations resulted in populations that were clearly not compatible with observations. These negative results provide, in some sense, as much useful information regarding planet formation models as the positive ones. 

\section{Comparison with observation}\label{sect:compresults}

\subsection{Statistical assessment}\label{subsection:statisticalassessment} 
In table \ref{tab:basicresutls}, the basic results for the nominal population \textit{i.e.} the population with the overall best results when compared to the actual population are summarized. The total number of initial conditions that were drawn is $\ninit=70\,000$. However, contrary to what may happen in real systems, we only start a formation calculation if the initial conditions are such that somewhere in the disk a sufficiently massive body ($\gtrsim \membstart=0.6$) can form during the disk's lifetime (see Paper I for an explanation).  

The later conditions is fulfilled in $\nsynt=50204$ ($\approx 72$ \%) of all disks. The corresponding $\nsynt$ planets constitute what we refer to as the ``full population'' despite the fact that it is incomplete at low masses (see next). In the remaining $\nnocalc=19796$ disks, a low $\fpg$ coincided with a low $\sigmanorm$ so that the isolation mass in the disk is $<0.6$ $\mearth$ everywhere, and/or the disk lifetimes is so short (high $\mwind$ together with a low $\sigmanorm$) that the disk disappears before such an embryo can form. For such disks which are hostile to planet formation in general and to giant planet formation in particular, we don't explicitly calculate the formation of a planet, as we have found that detectable synthetic planets can form only if the disk lifetime is at least $\sim0.5$ Myr (\S \ref{subsect:formation timescales}) and the isolation mass is larger than about 3 Earth masses (usually it is of the order of $8$ $\mearth$ or larger).  Therefore, detectable synthetic planets cannot form in the $\nnocalc$ disks. Thus, the fact that $\nnocalc$ disks/initial conditions are discarded has no influence on the statistical analysis of the detectable sub-populations which is our primary interest in this paper. 

Qualitatively we expect that in the $\nnocalc$ disks planets will form also. Most likely, a system of very low mass planets (less than a few $\mearth$) will eventually emerge. However, as their formation is likely to occur on timescales significantly longer than the gas disk lifetime, these planets will not be able to accrete nebular gas. The resulting incompleteness of our model at low masses should be kept in mind when considering our predictions regarding the full population, for example in the initial mass function (fig. \ref{fig:imf}). 

Within the  $\tmc=10$ yrs survey length and with the $\epsilonmc=10$ m/s instrumental precision, 6075 planets with $a>\atouch$ are classified as detectable in our synthetic RV survey.  $\nhot$=1386 planets migrated to $\atouch$. We call such cases ``Hot'' planets (see \S\ref{subsect:obsconsthotjup}). Assuming that all $\nhot$ planets were swallowed by their host star, the synthetic survey has thus an overall detection probability of $P=6075/70\,000=8.7$ \%. Assuming the other extreme case, namely that all $\nhot$ planets became detectable (regardless of their actual mass when they reach the ``feeding limit'' at inner boarder of our computational disk at $\atouch\approx 0.1$ AU), leads to a  $\pwhot=(6075+1386)/70\,000=10.7$ \%. These two values could be seen as bracketing the value for the real detection probability unless the effect of the one-embryo-per-disk approach changes the picture too dramatically (cf. \S\ref{subsect:definingobsconstarints}). 

Nevertheless, we notice that the overall detection probability from the synthetic survey agrees surprisingly well with often quoted actual yields of 5-10 \% as for example the 6.6\% given by Marcy et al. (\cite{marcyetal2005}) for giant planets with $a\lesssim$5 AU as in our simulation. Cumming et al. (\cite{cummingetal2008}) find an extrapolated occurrence rate of planets with $0.3<\msini/\mj<15$ of $8.5\pm1.3$ or $11\pm1.7 \%$ out to a semimajor axis of 3 and 5 AU, respectively.

For the KS tests, the number of bootstrap samples $\nsamp$ is 180. The specific KS results for $a-\msini$, $\msini$, $a$ and [Fe/H] are discussed below. 

\begin{table}
\begin{center}
\caption{Basic results of the population synthesis.}\label{tab:basicresutls}
\begin{tabular}{ll}\hline
Feature & Value\\\hline
Duration of synthetic survey ($\tmc$) [yr] &  10 \\
RV-precision of synthetic survey ($\epsilonmc$) [m/s] &  10 \\ 
Nb. of initial conditions ($\ninit$) & 70\,000 \\
Nb. of calculated synthetic planets ($\nsynt$) & 50\,204 \\
Nb. of initial conditions without calculations ($\nnocalc$) & 19\,796 \\ 
Nb. of detectable synthetic planets ($\nobssynt$) & 6075 \\
Nb. of synth. planets migrating to $\atouch$ ($\nhot$) &  1386 \\
Synth. detection probability w/o $\nhot$ ($P$) [\%] & 8.7 \\ 
Synth. detection probability w. $\nhot$ ($\pwhot$) [\%] & 10.7 \\ 
Nb. of actual planets in obs. comp. sample ($\nobsreal$) & 32 \\
Nb. of KS bootstrap samples ($\nsamp$) & 180 \\
Significance KS $a-\msini$ ($\sam$)  [\%] & 87.7 \\
Significance KS $\msini$ ($\sm$) [\%] &  95.6 \\
Significance KS $a$ ($\sa$) [\%] & 63.9 \\
Significance KS [Fe/H] ($\sfeh$) [\%] & 21.7  \\
\hline
\end{tabular}
\end{center}
\end{table}

\subsection{Mass-Distance diagram}\label{subsect:aMdiagram}
The KS test for the two dimensional distribution in the mass-distance plane actually checks wether planets of the correct mass are found at the correct distance. It is the observational constraint that we weighted highest, as the mass-distance diagram is of similar importance for planet formation and evolution as the Hertzsprung-Russell diagram for stars (Ida \& Lin \cite{idalin2004a}). The reason for this  is that it contains a lot of information about the planetary formation process, as described in Paper I.

When comparing our synthetic results with the observed distribution, one should bear in mind that our model shows the mass-distance distribution at the time when the gaseous disk disappears. Later on, it can be modified by evolutionary effects as evaporation or N-body interactions in initially more crowded systems. For the statistical comparison, we have tried to minimize those effects by carefully choosing the observational comparison sample, as described in \S\ref{subsection:observationalcomparisonsample}.

\subsubsection{Full population}\label{subsubsect:aMfullpopulation}
\begin{figure*}
	\centering
      \includegraphics[width=19cm]{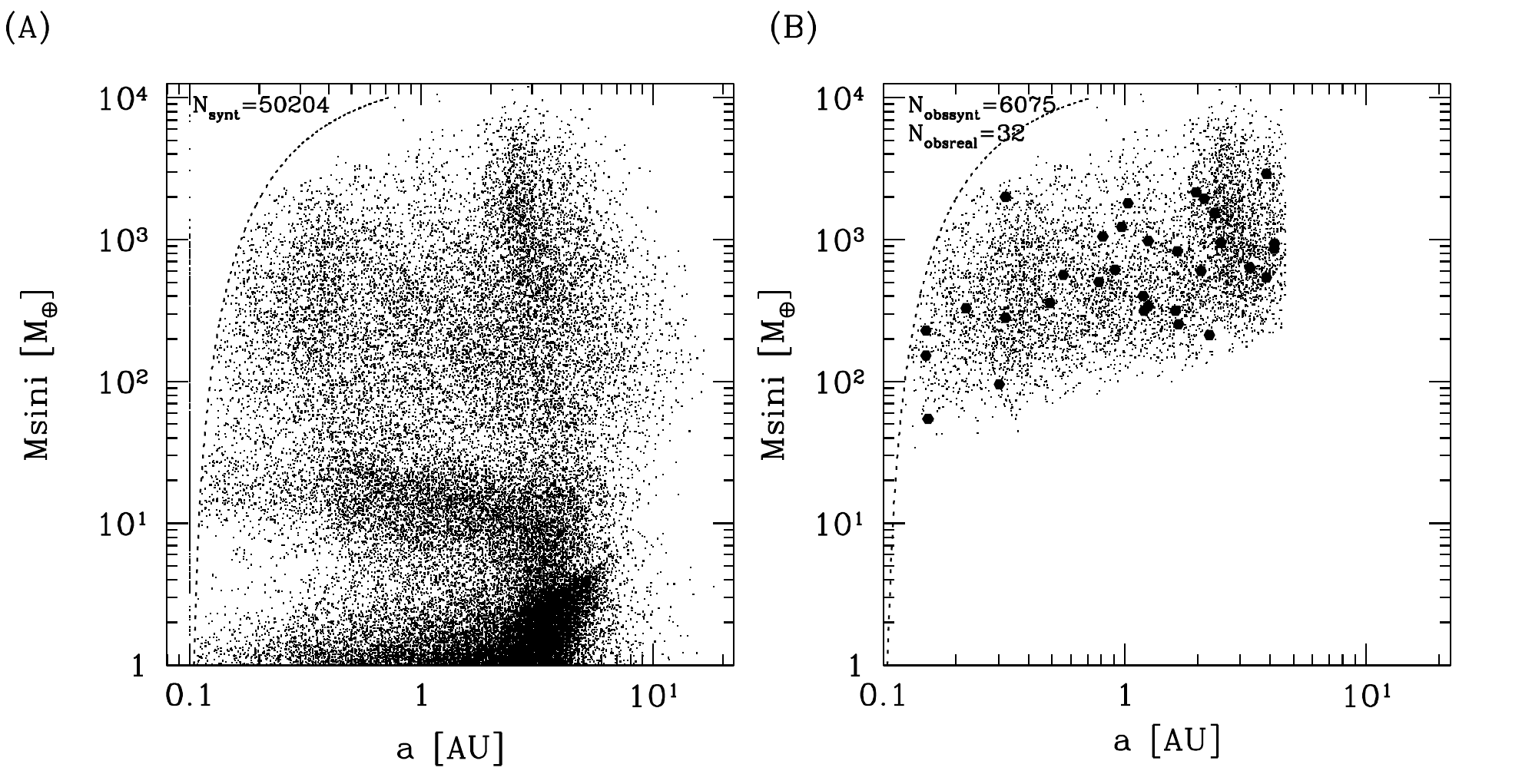}
      \caption{\textbf{Panel (A)}: Projected mass $\msini$ versus distance $a$ of the full synthetic population. The feeding limit at $\atouch$ is plotted as dotted line. Planets migrating into the feeding limit have been put to 0.1 AU. Various sub-populations which have been discussed in Paper I can be distinguished, namely the ``failed cores'', the  ``horizontal branch'', the ``main clump'', and the ``outer group''. As expected, the effect of $\sin i$ is to blur slightly these structures. \textbf{Panel (B)}: The remaining sub-population of $\nobssynt=6075$ actually detectable synthetic planets after applying the $\epsilonmc=10$ m/s synthetic RV bias (small dots). The sharp cutoff at about 4.6 AU corresponds  to an orbital period of 10 years, the assumed observational baseline $\tmc$.  The observational comparison sample with $\nobsreal=32$ real extrasolar planets is overplotted as big dots.} 
      \label{fig:amdist}
\end{figure*}

In fig. \ref{fig:amdist}, panel (A), the projected mass versus distance diagram of the full synthetic population is plotted. In the companion Paper I, we have thoroughly discussed this figure, so that we only summarize these findings here. 

One first notes that the variation of initial conditions within the observed boundaries results in a synthetic planet population with a large diversity.  We conclude that the observed diversity of extrasolar planets is  a natural consequence of the diversity of protoplanetary disks. Inside the envelope covered by the planets, various sub-structures like concentration, clumps, bars or depletions can be identified. 

The most prominent concentration can be seen near the lower mass boundary. There is a vast sub-population of  core dominated low mass planets ($M\approx\miso\lesssim5-10$ $\mearth$). We call this group the  \textit{``failed cores''} in the sense that these planets did not manage to grow large enough within the lifetime of the gaseous disk to accrete significant amount of gas. These ``failed core'' planets are however not identical to the final population of terrestrial planets. They rather represent an earlier state in the formation of terrestrial planets and more massive icy counterparts beyond the iceline. That is to say they show the state of such planets at the moment when the gas disk disappears. Growth beyond isolation where terrestrial planets obtain their final mass occurs through a series of giant impacts among protoplanets of comparable size (which can be identified with the ``failed cores'') after the gaseous disk is gone in a final, post-oligarchic phase (Goldreich et al. \cite{goldreichetal2004}).  Then, all ``failed cores'' in one disk (of which we however model only one) start to interact gravitationally, leading to a rearrangement through scattering and ejections and to mass growth by giant impacts (Ford \& Chiang \cite{fordchiang2007}). Such a behavior can be seen in the simulations of Thommes et al. (\cite{thommesetal2008}), where planetary systems with a number (of order 10) low mass planets are the typical simulation outcome for low mass disks.

At semimajor axes between the feeding limit at roughly 0.1 AU and approximately 5 AU, and masses of $10\lesssim M/\mearth\leq30$, a  \textit{``horizontal branch''} of subcritical  ($\menv/\mheavy\lesssim 1$) cores is visible. Their seeds started usually outside the iceline. While some planets are found at intermediate semimajor axes at the moment when the disk disappears,  the ``horizontal branch'' also also acts as the ``conveyor belt'' by which many low mass planets are transported close to the star ($<0.1$ AU).  Observed examples of the ``horizontal branch'' might be the Neptune-mass planets around \object{HD 69830} (Lovis et al. \cite{lovisetal2006}). 

While migrating along the ``horizontal branch'', some planets become supercritical for gas runaway accretion and leave the branch upwards towards higher masses (cf. the formation tracks in Paper I).  This leads to a concentration of giant gaseous planets at distances from the star of roughly 0.3 to 2 AU and masses of  $100\lesssim M/\mearth\lesssim 1000$. We call this concentration the \textit{``main clump''}. 

We find that once gas runaway is triggered, the planetary gas accretion rate becomes quickly limited by the gas accretion rate in the disk and the planetary mass grows larger than the local disk mass. As explained in Paper I, limiting the planetary gas accretion rate by the disk accretion rate has the consequence that we find a clearly less pronounced depletion of planets of intermediate masses than it is the case for the \textit{``planetary desert''} found first by Ida \& Lin (\cite{idalin2004a}). A certain depletion of planets with masses between $\sim30-100$ $\mearth$ is however visible in the mass-orbit diagram of our synthetic population also as well as in the planetary IMF, fig. \ref{fig:imf}, as it is a typical feature of core accretion. The second effect (a higher planetary mass than the local disk gas mass), has as consequence that the migration mode changes from the disk dominated regime into the slower planet dominated type II mode  (Paper I; Edgar \cite{edgar2007}).  This slowing down prevents the planets from migrating too quickly towards the star. It also naturally causes an absence of very massive planets at small distances from the star, as observed (Zucker \& Mazeh \cite{zuckermazeh2002}).

Planetary seeds with a large starting position $\astart$ (between 4-7 and 20 AU),  and a disk environment with a high solid surface density (high $\fpg$ and/or $\sigmanorm$) can grow supercritical for gas runaway accretion in-situ and become giant planets, without the need of passing first though the ``horizontal branch'' to collect enough solids.  This leads to the formation of another concentration of giant planets, the \textit{``outer group''} with $2 \lesssim a \lesssim 5$ AU and $1 \lesssim M/ \mj \lesssim 20$. Some of the planets in this group are thus so massive that they fall in the interesting category of deuterium burning planets (Baraffe et al. \cite{baraffechabrierinprep}).

The mass-orbit diagram also shows a second depletion of planets at semimajor axes inside 2 AU between the ``failed core'' planets and those in the ``horizontal branch'', \textit{i.e.} at masses between $\sim3$ and 10 $\mearth$. At small semimajor axes ($\lesssim 0.3$ AU) it is particularly clear. The reason is the following: ``Failed cores'' which only migrate in the strongly reduced type I mode in the nominal case, grow only up to a mass approximately equal the isolation mass $\miso$. For the most metal rich disks considered here ($\sigmanorm=1000$ g/cm$^2$, $\fpg=0.13$), $\miso$ at 0.1 AU is about 3 $\mearth$. Planets in the ``horizontal branch'' have in contrast  accreted most of the solids in the inner part of the disk once the reach small semimajor axes, so that they have a minimal mass of the order of 10 $\mearth$ at 0.1 AU. Note that the faster type I migration, the lower this mass limit (Paper I, \S \ref{subsubsect:f1variation}).  Growth beyond the isolation mass up to final masses by giant impacts between different ``failed cores'' would tend to fill the depleted region. We can roughly estimate the mass to which Super Earth planets could grow by this process in-situ. If all originally present solids are incorporated into the planets, and their final relative spacing is of the order of $\Delta a \sim a/3$ as in the solar system (Goldreich et al. \cite{goldreichetal2004}), then planets as massive as 10 Earth masses could form at 0.1 AU in the most metal rich disk which would fill up at least partially the depleted region.    

This second depletion, which is also in visible in the planetary mass spectrum (\S \ref{subsubsect:planetaryIMF}) is therefore not a robust prediction of the model, and could in principle disappear once planet growth after disk dispersion is included in the model. Note that for the statistical comparison of the detectable synthetic planets with our comparison sample of actual known exoplanets, this does not constitute an issue.   

\subsubsection{Detectable sub-population}
Panel (B) in fig. \ref{fig:amdist} shows the sub-population of the potentially \textit{detectable} synthetic planets which remains after applying the $\epsilonmc=10$ m/s synthetic bias of \S\ref{subsection:syntheticRVbias}.  The sharp cutoff at about 4.6 AU corresponds to a 10 year period, the assumed observational baseline $\tmc$. The observational comparison sample with $\nobsreal=32$ real extrasolar planets is overplotted as big dots. The most striking feature is that our synthetic MC survey is able to detect just a small fraction of the underlying full planet population, between 8.7\% to 10.7\% of all synthetic planets (tab. \ref{tab:basicresutls}). As expected, the planets detectable at 10 m/s are Saturn to Super Jupiter class planets, plus a handful planets with intermediate masses ($\sim50$ $\mearth$) close to the star. Even if radial velocities measurement in the last few years have reached a precision much better than 10 m/s (see \S\ref{subsect:obspopulationRV} for the detectable sub-population at $\epsilonmc=1$ or 0.1 m/s), we still can conclude that the currently known extrasolar planets are just the tip of the iceberg of the real underlying population, as discoveries at the 1 m/s level still require a large investment of observational time and are restricted to small semimajor axes. At the high mass end, we see that the ``outer group'' represents a significant reservoir of very massive planets at larger semimajor axes. We note that Cumming et al. (\cite{cummingetal2008}, their fig. 5) have shown that in the Keck Planet  search program a group of very massive candidates ($\msini\gtrsim20$ $\mj$) at periods $\gtrsim2000$ d ($a\gtrsim 3$ AU) exists which have not yet been announced. Such very massive objects are virtually absent in the model at smaller semimajor axes (and especially do not reach the feeding limit), in agreement with observations (Udry et al. \cite{udryetal2003}). 

In the statistical comparison of the detectable sub-population with the observational comparison sample we find with the two dimensional KS test of the $a-\msini$ distribution a significance of 87.7\% that the two populations come from the same parent population. To our knowledge, this is the first time that it is shown that a theoretical formation model is able to reproduce in a quantitative way the observed mass-distance distribution of an adequate comparison sample of extrasolar giant planets.  

Even if we have determined the two most important parameters of the model, $\alpha$ and $\f1$ by fitting the detected planet population, getting an agreement for any combination of parameters is a nontrivial result. First, a certain number of elements were given such as the formation model itself and the probability distributions for $\fpg$, $\sigmanorm$, and $\mwind$ which were derived from observations. Second, the number of observational constraints that must be reproduced \textit{concurrently} with one population is large (\S \ref{subsect:definingobsconstarints}), while the number of free parameters is small. Third, at least one of the parameters, $\alpha$, can only be varied within about one order of magnitude as observational constraints exist (King et al. \cite{kingetal2007}).  Finally, varying parameters has complex consequences on the characteristics of the population thereby limiting the possibility to force the system in a particular direction. We therefore interpret this result together with the other ones of this section as an indication that the core accretion mechanism as described here, while still being extremely rudimentary, must successfully catch several essential aspects of giant planet formation. 

\subsection{Mass $\msini$}\label{subsect:msini}
The second distribution we have compared statistically is the mass distribution. It is clear that good results in the 2D $a-\msini$ distribution imply to some extent good results for the 1D distributions of $\msini$ and $a$ separately (whereas the opposite is not true). It is nevertheless worth studying these important distributions also separately, as they have been discussed extensively from both an observational and theoretical point of view (e.g. Udry \& Santos \cite{udrysantos2007}; Ida \& Lin \cite{idalin2004b}), and because it is simpler to gain in this way insights into the differences between model and observation than in the 2D case.  

\subsubsection{Planetary IMF}\label{subsubsect:planetaryIMF}
\begin{figure}
      \centering
  \resizebox{\hsize}{!}{\includegraphics{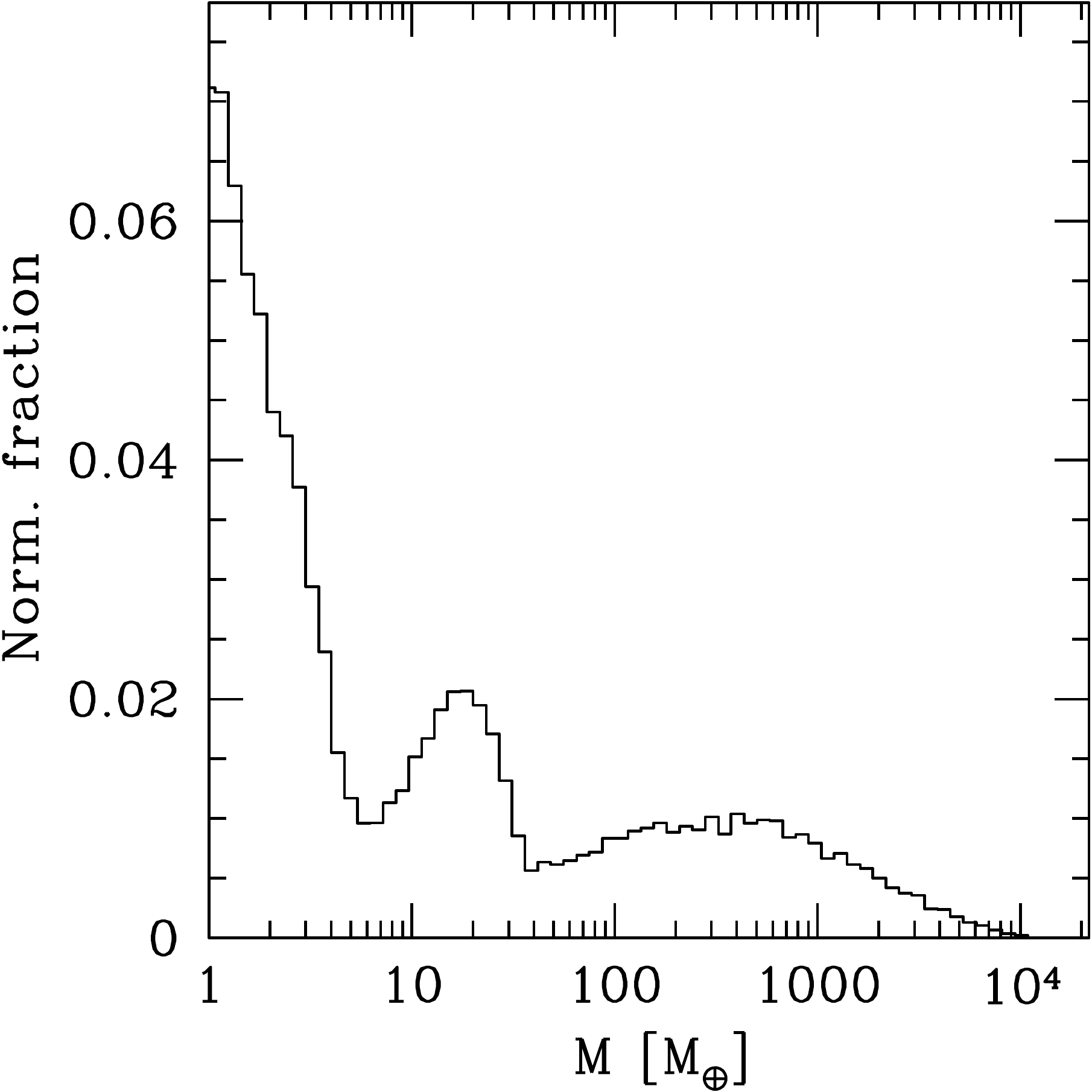}}
      \caption{Planetary initial mass function, corresponding to the moment in time when the gaseous protoplanetary disk disappears. Several mechanisms can subsequently modify the distribution. The largest changes are expected to occur at low masses (below $\sim 10-20$ $\mearth$). The planetary IMF (or PIMF) has a complex structure with several minima and maxima (see text).} 
      \label{fig:imf}
\end{figure}

Before comparing the detectable sub-population with the real observations, it is interesting to have a look at the underlying, unbiased mass distribution of the full synthetic population, as this can have important implications for future planet search campaigns. Figure \ref{fig:imf} shows the predicted planetary initial mass function (PIMF) of synthetic planets around 1 $\msun$ stars in the one-embryo-per-disk approximation. One should keep in mind that this PIMF shows the planetary mass spectrum at the moment when the disk disappears. Subsequent modifications due to evaporation, planetary merging and ejection by N-body interactions and especially the formation of low mass planets on long timescales are not included in our model. 

At the largest masses, the PIMF shows a smooth decrease with increasing mass through the Super Jupiter ($M \gtrsim 3$  $\mj$) and the Deuterium Burning Planets ($M \gtrsim 13.6$ $\mj$, cf. below in this section) mass domain. While this tail contains only a small percentage of the full population (tab. \ref{tab:planettypesbymassfractions}), it is obviously of primary interests for several detection techniques such as astrometry or direct imaging. Planets in this domain are mainly formed in disks with a lifetime above average (high $\sigmanorm$ and low $\mwind$), in which the planet has plenty of time to accrete all nebular gas that is not photoevaporated and that viscously flows towards its position.

At lower masses, in the Jovian mass domain ($100\lesssim M/ \mearth\lesssim 1000$), the mass distribution is approximatively flat, although a shallow local maximum near 1-2 Jupiter mass can be seen, as it is observed (e.g. Jiang et al. \cite{jiangetal}). 

Decreasing further in mass to intermediate sized planets with no equivalence in the Solar System ($30\lesssim M/\mearth \lesssim 100$), we see a smooth decrease of the PIMF. It falls to a local minimum at about 30-40 $\mearth$. This minimum is a consequence of the same effect that causes the ``planetary desert'' (Ida \& Lin \cite{idalin2004a}), \textit{i.e.} that fact that once gas runaway starts, it quickly leads to a significant increase in mass, so that the probability for a planet to have a final mass just slightly larger than the one needed for runaway accretion is small because it is unlikely that disk dispersion cuts the gas supply exactly at this moment. This minimum can therefore be seen as the transition between low mass, solid dominated planets, and gas dominated giant planets.

It is interesting to note that this effect, which is very characteristic for the core accretion mechanism is well visible as minimum in the PIMF. Its location at a total planetary mass of about 30 Earth masses is expected, as this is the typical total mass when gas runaway accretion sets in near the crossover mass, \textit{i.e.} when $\mcore\approx\menv\approx15$ $\mearth$ (e.g. Pollack et al. \cite{pollacketal1996}). An observational confirmation of the minimum at about 30 $\mearth$ (which becomes visible at a RV precision of 1 m/s, see fig. \ref{fig:msinihisto3PRV}) indeed seems to have occurred very recently  (Mayor \& Udry \cite{mayorudry2008}). This would be a strong indication in favor of core accretion as the dominant giant planet formation channel, and its location would be a direct measurement of the mass at which runaway accretion typically begins. 

Note that the intermediate mass domain (30-100 $\mearth$) is still quite populated in our calculations (the PIMF shows that there are about 2.5 times less planets in the intermediate mass than in the Jovian mass domain, see tab. \ref{tab:planettypesbymassfractions}). Radial velocity surveys only now start to carry out observations with the precision required to  detect planets in this mass range at a significant fraction of an AU  (Lovis et al. \cite{lovisetal2006}).  High precision RV measurements over a long time baseline will be very helpful in observationally characterizing this part of the PIMF from which the characteristic timescale for runaway gas accretion could be derived since the later determines how populated this intermediate mass range will be. This would in turn provide important information for formation models. For example have Miguel \& Brunini (\cite{miguelbrunini2008}) recently shown that they find a minimum in the planetary mass spectrum at 100-1000 $\mearth$ if they use a formation model similar to Ida \& Lin (\cite{idalin2004a}) but couple it to solid and gas accretion rates which fit the results of Fortier et al. (\cite{fortieretal2007}), rather than a minimum at 10-100 $\mearth$ as in Ida \& Lin (\cite{idalin2004a}).

At masses below $30-40$ $\mearth$, the PIMF raises rapidly again with decreasing mass to reach a well defined local maximum in the Neptunian mass domain (7-30 $\mearth$). The maximum at about 15 Earth masses is caused by the planets in the ``horizontal branch'' in the $a-\msini$ diagram, \textit{i.e.} of subcritical cores migrating in disk dominated type II and  collecting solids (Paper I). These planets have properties similar to the ice giants of our own planetary system with $0.01\lesssim\menv/\mcore\lesssim0.4$.

A next local minimum occurs at about $7$ $\mearth$. It marks the boundary between ``failed cores'' and planets in the ``horizontal branch''. The reason for this minimum is clear as well. In order to best reproduce the observed population of giant extrasolar planets, especially the semimajor axis distribution, we found that very small type I  migration efficiency factors are needed (see section \ref{subsubsect:f1variation}). With such low type I migration rates, isolation phenomena in which the planet almost completely empties its feeding zone of solids at its starting position become important. This results in the quenching of the embryo's growth at a mass of roughly  $\lesssim 5$ $ \mearth$, which approximately corresponds to the maximal mass to which a ``failed core'' can grow in a disk with a mean solid surface density just beyond the iceline. Type II migration is in contrast calculated at its nominal rate. Therefore, as soon as the migration mode switches from type I to type II, the core moves quickly into new, undisturbed regions of the disk with a large supply of new planetesimals to accrete and therefore resumes its growth to reach masses of order $M\sim 20$ $\mearth$ which is the typical mass an embryo reaches  after it has swept significant parts of the inner regions of the disk while migrating through the ``horizontal branch''. Hence, except for unlikely timing effects, embryos with a final mass between the two values mentioned above are not as frequent as the others. 

Finally, the PIMF begins to raise very quickly in the Super Earth mass domain, indicative of the large population of low mass ``failed cores'' planets already seen in the mass-distance diagram.   

\begin{table}[t]
\begin{center}
\caption{Percentage of various planet types forming from the $\ninit=70\,000$ initial conditions that were generated (``Init''), of the detectable synthetic sub-population (``Obssynt'', $\nobssynt=6075$ planets), and of the observational comparison sample (``Obsreal'', $\nobsreal=32$ planets). For the latter, Poisson errors are assumed. For the full population, the true mass $M$ is used, for the other two populations the projected mass $\msini$ to allow comparison.  ``Not calculated'' corresponds to the $\nnocalc=19796$ initial conditions where no 0.6 $\mearth$ seed could be formed during the disk lifetime. DBP stands for Deuterium Burning Planets.}\label{tab:planettypesbymassfractions}
\begin{tabular}{lllll}\hline
Type [\%] & Range [$\mearth$]  & Init    & Obssynt & Obsreal \\ \hline
Not calculated  &  -                & 28.3  & -       & -\\
Super Earth      &$ <7$          & 41.4  &0       & 0\\
Neptunian          & 7-30          & 11.8  &0        &0   \\
Intermediate      & 30-100      & 4.2    & 2.4   & $6.3\pm4.4$\\
Jovian                & 100-1000   & 10.4 & 66.6 & $68.8\pm14.7$\\
Super Jupiter   & 1000-4323 &  3.5  &28.1  & $24.9\pm8.8$\\
DBP                     & $>$4323    & 0.4   &2.9    & 0\\ \hline
\end{tabular} 
\end{center}
\end{table}

\begin{figure*}
      \centering
      \includegraphics[width=19cm]{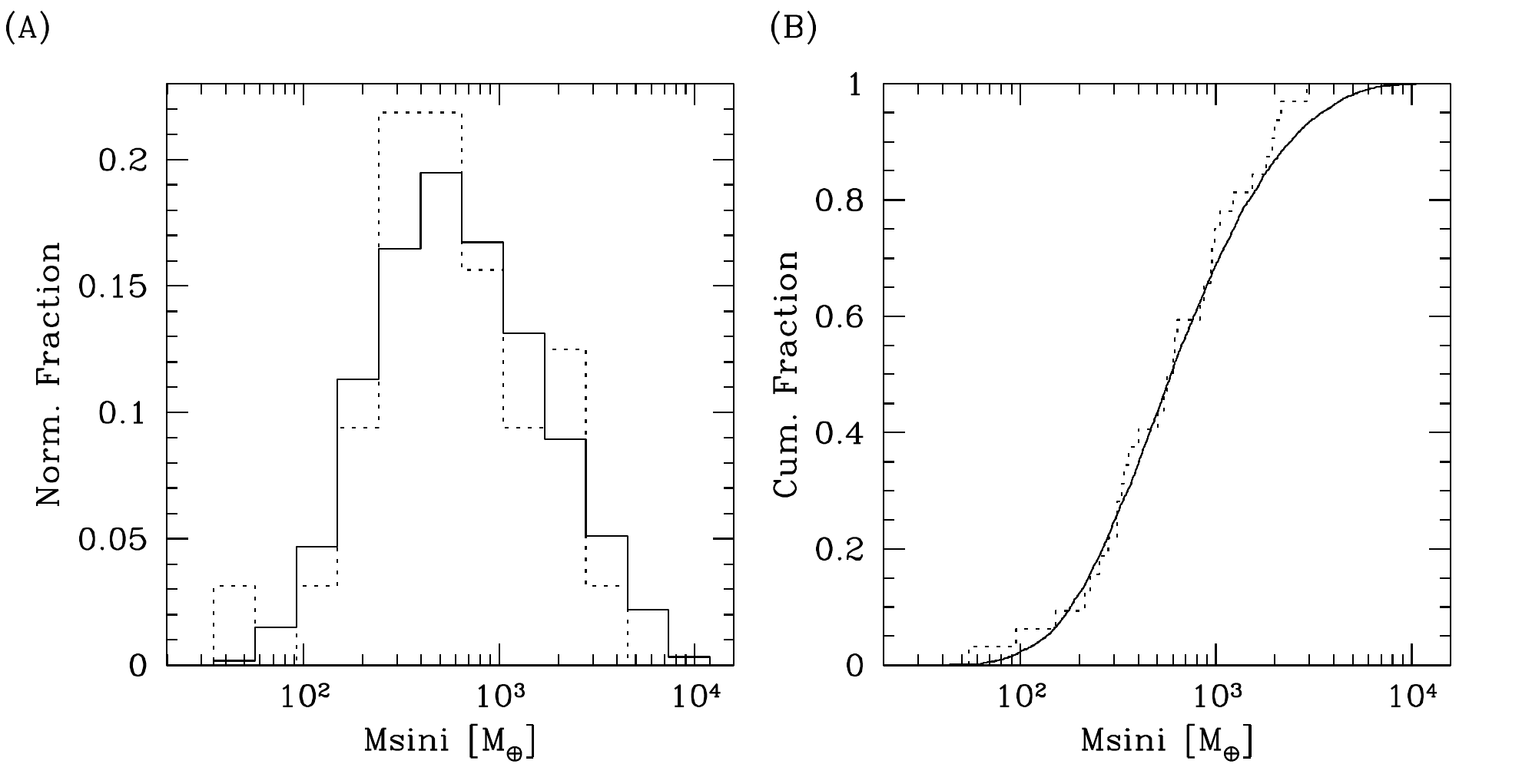}
      \caption{Statistical test of the mass distribution. \textbf{Panel (A)}:  Histogram of the projected mass $\msini$ of the sub-population of the $\nobssynt=6075$ detectable synthetic planets (solid line) and the observational comparison sample of 
      $\nobsreal=32$ planets (dotted line).
      \textbf{Panel (B)}:  Cumulative distribution function corresponding to (A), showing that the two distributions are very similar.}
       \label{fig:msinidist}
\end{figure*}

The quantitative characteristics of the planetary IMF at such low masses (\textit{i.e.} below 10-20 Earth masses) predicted by our model should be considered with great caution. As we have already mentioned, is our model essentially a giant planet formation model and is therefore severely incomplete at the low mass end (see also Paper I). For example, the minimum at 7 $\mearth$ between Super Earth and Neptunian planets might be an artifact resulting from not having considered the merging of ``failed core'' after disk dispersal as well as planetary formation in the $\nnocalc$ solid-poor disks.  Despite this caution, we nevertheless believe that a strong rise of the PIMF somewhere below $\sim10-20$ $\mearth$ is a real feature simply reflecting the fact that the vast majority of embryos are located in disks that do not allow the formation of massive cores.

Forming Jovian planets is in contrast only possible when initial conditions are, by chance, planet formation friendly which corresponds to a disk with significant amounts of solids and gas. It is interesting to note in this context that the Sun is in the low metallicity tail of planet host stars (Udry \& Santos \cite{udrysantos2007}). Under these favorable conditions, the tip of the iceberg of the full planetary population is formed, the one we detect with radial velocity techniques at a precision of 10 m/s. Usually, our attention is focussed on stars which do have detectable (giant) planets. But the finding that 90\% of the stars in RV surveys do not appear to have detectable giant planets is an important result as well and is in full agreement with our PIMF.

From the consideration of the PIMF, we compute the fractions of different types of synthetic planets in table \ref{tab:planettypesbymassfractions}. The numbers for the lowest mass bins must very likely be seen as lower limits. For example, many of the $\nnocalc$ initial conditions for which we did not calculate the formation of a planet as no 0.6 $\mearth$ seed could be formed during the lifetime of the gas disk, may eventually lead to the  formation of  planets with masses below 7 $\mearth$. If this were to be the case, this mass bin would eventually contain 69.7 \% of all initial conditions. 

It is clear that the exact value of the masses defining the various bins in tab. \ref{tab:planettypesbymassfractions} are somewhat arbitrary but are nevertheless based on the various features of the PIMF. In any case it is interesting to note that the maxima of the PIMF correspond roughly to planetary masses occurring in our solar system. 

Finally, one notes that the core accretion mechanism is able (under certain assumptions, see Paper I) to produce planets which lie in a mass domain where, at least in the absence of a solid core, deuterium fusion occurs ($M\gtrsim0.012-0.013$ $\msun\approx12.6-13.6$ $\mj$, Chabrier \& Baraffe \cite{chabrierbaraffe2000}). Therefore, we simply call all planets with a mass larger than 13.6 $\mj$ ``Deuterium Burning Planets'' (DBP, see Baraffe et al. \cite{baraffechabrierinprep}). Such massive planets are however rare objects: only 0.4 \% of all disks produce a planet larger than 13.6 $\mj$, and just 11 out of  70\,000 initial conditions produce an object more massive than 30 $\mj$. The existence of the ``brown dwarf desert''  is therefore not in disagreement with these findings. For example, Marcy \& Butler (\cite{marcybutler2000}) have estimated a frequency of brown dwarfs within 3 AU of $\lesssim0.5\%$. The overall largest synthetic planet has a mass of 38 $\mj$, which is just in the middle of the desert (Lovis \& Mayor \cite{lovismayor2007}).

\subsubsection{Observed mass distribution}
In fig. \ref{fig:msinidist}, the two panels illustrate the results of the statistical comparison of the $\msini$ distribution of the detectable sub-populations with the observational comparison sample. Panel (A) of fig. \ref{fig:msinidist} shows the mass histogram of the sub-population of the detectable synthetic planets, together with the observational comparison sample. Both distributions peak at about 2 $\mj$. The decrease at high masses simply reflects the decrease in the unbiased PIMF in the Super Jupiter and DBP domain. The decrease at low masses is mainly an effect of the observational detection limit of 10 m/s, but not entirely, because also the underlying distribution decreases below $\sim1$ $\mj$.

Panel (B) shows the corresponding cumulative distribution functions. The plots make it obvious that the two distributions are very similar. Some differences exists at the low and high mass end. At the high mass end this could be related to the fact that the $\nobsreal=32$ observational comparison sample only includes companions if their mass is smaller than 20 $\mj$, while we do not use such a criterion, so that we might should add some brown dwarf candidates, as suggested by Lovis \& Mayor (\cite{lovismayor2007}). 

Applying the KS test results in a high significance that both the observed and the synthetic population are drawn from the same parent distribution. The bootstrap method leads to $\sm=95.6$\%, while using eqs. \ref{eq:significanceKSfromdKS1} to \ref{eq:significanceKSfromdKS3} leads to 96.8 \%. We have found that it is possible to obtain good, or at least non-zero statistical significance for the mass distribution even if some parameters were changed (cf. \S\ref{subsect:influeceofparameters}), in contrast to $\sam$ or $\sa$, that often fall to virtually zero. These surprisingly good results for the mass distribution has led us to conclude that gas accretion by a core may be relatively well understood as compared to, for example,  the migration of it. 

\begin{figure}
   \resizebox{\hsize}{!}{\includegraphics{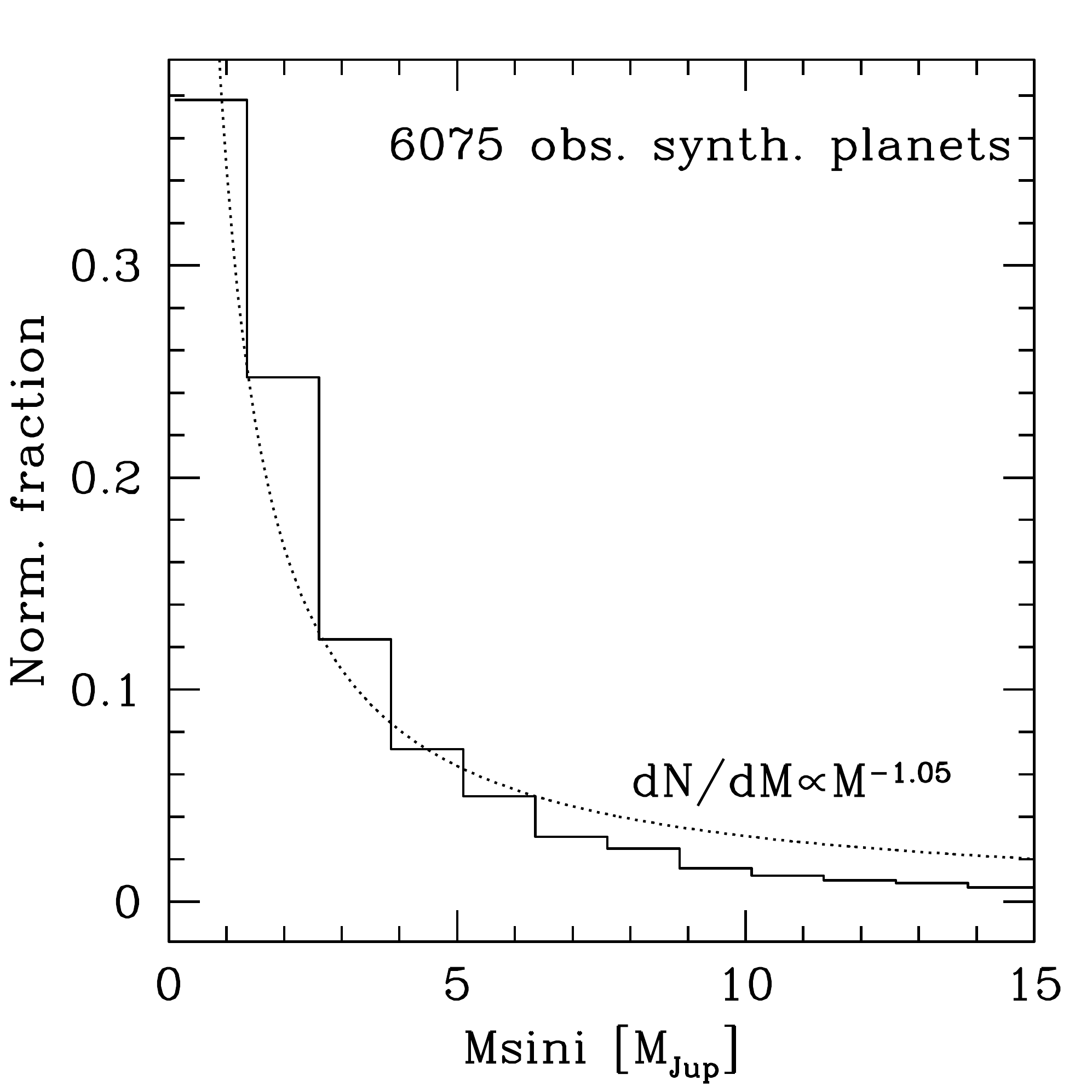}}
   \caption{Histogram of $\msini$ of the 6075 detectable synthetic planets compared to the observationally inferred d$N$/d$M\propto M^{-1.05}$ power law of Marcy et al. (\cite{marcyetal2005}). }  
   \label{fig:mhistomarcy}
\end{figure}

The  good agreement with observational data is also visible in fig. \ref{fig:mhistomarcy}, where the mass distribution of the detectable sub-population is compared in linear bins to Marcy's et al. (\cite{marcyetal2005}) power law d$N$/d$M\propto M^{-1.05}$ inferred from the Keck, Lick \& AAT observational data (See Udry \& Santos \cite{udrysantos2007} for an updated version of this plot.). Similarly to the observational result this power law gives a good representation of the mass distribution in the Jovian planet regime. This is due to the approximatively flat part (in logarithmic units) of the PIMF in the Jovian mass domain discussed above.  At masses much larger than one Jupiter mass ($\gtrsim7$ $\mj$), and especially in the DBP domain, the number of planets is clearly lower than inferred from the d$N$/d$M\propto M^{-1.05}$ law, again in good agreement with Marcy et al. (\cite{marcyetal2005}) or Butler et al. (\cite{butleretal2006}).  This is due to the decrease of the PIMF in the Super Jupiter and DBP tail. From the complex shape of the unbiased PIMF in fig. \ref{fig:imf} it is however also clear that one power law cannot be used to describe the full domain of planetary masses as already noted by Mayor \& Udry (\cite{mayorudry2008}).

\subsection{Semi-major axis $a$}\label{subsect:semimajoraxis}
\begin{figure*}
\centering
      \includegraphics[width=19cm]{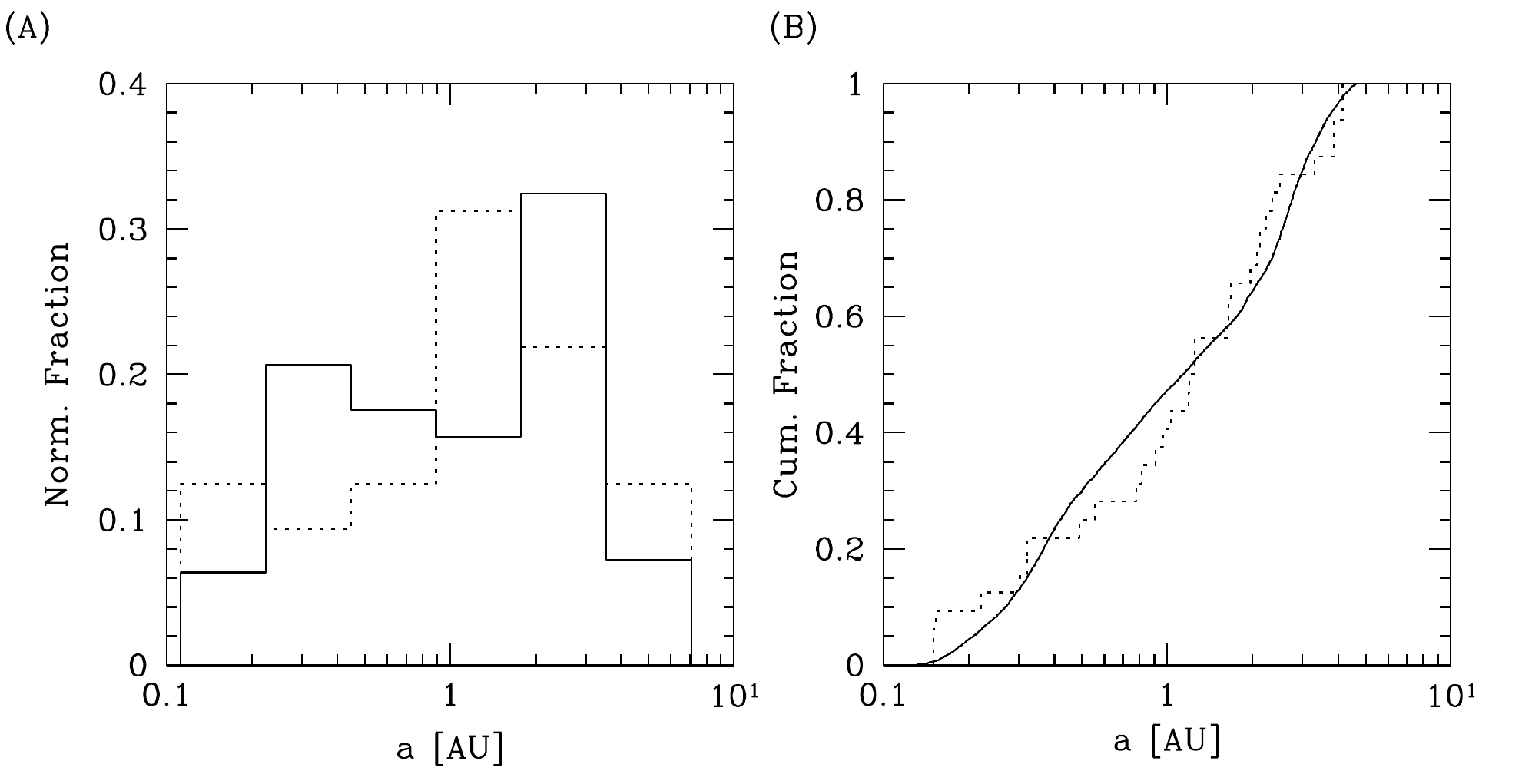}
      \caption{Statistical test of the semi-major axis distribution.
       \textbf{Panel (A)}: Distribution of the final semimajor axes of the detectable synthetic planets outside the feeding limit (solid line) and the observational comparison sample (dotted line).   
      \textbf{Panel (B)}: Cumulative distribution function corresponding to (A).}
      \label{fig:adist}
\end{figure*}

The third distribution we have compared with the observations is the semimajor axis.  In the comparison, the  ``Hot'' synthetic planets that reach the inner boundary of the computational disk at $\atouch$ are again not included, as the structure of the disk closer than $\sim0.1$ AU to the star is likely to be more complicated than described in the model, and probably causes particular effects, like stopping planetary migration. Correspondingly, observed Hot Jupiters are neither included (\S \ref{subsection:observationalcomparisonsample}). The constraints these planets put on the model are treated in \S\ref{subsect:obsconsthotjup}.    

In fig. \ref{fig:adist}, Panel (A) the distributions of the semimajor axes of  the detectable synthetic planets and the $\nobsreal=32$ observational comparison sample is plotted. Although the two histograms are similar (an approximately flat distribution at smaller distances followed by a sharp upturn at larger distances),  two  differences exist. First, among the synthetic population there are more detectable planets inside $\sim 1$ AU than in the observational sample, and second, the upturn occurs in the model at a larger distance ($\sim 2$ AU) instead of about 1 AU as in the observational comparison sample, or as e.g. also shown in Marcy et al. (\cite{marcyetal2005}).

In the model, the inner flat part is populated by giant planets of the ``main clump'' which first migrate through the ``horizontal branch'' to collect solids (Paper I), while the upturn at $\sim 2$ AU is caused by ``outer group'' planets which grow supercritical for gas runaway accretion in-situ. 

Giant planets at  smaller distances thus have a different, more complicated formation and especially migration history, than those further away. The number of giant planets inside $\sim 1$ AU is thus dependent on the efficiency with which migrating cores can accrete planetesimals, or the solid surface density profile below 1 AU. Both these factors are described only roughly in the model as the accretion rate is the same for a migrating as for a non migrating planet, and we use a simple $\propto r^{-3/2}$ law for the solid surface density, which could be modified for example by planetesimal drift due to gas drag which occurs on the fastest timescales at small distances (e.g. Chambers \cite{chambers2006}). The location of the ``outer group'' and thus the upturn at about 2 AU is dependent on the location of the iceline. In tests where we have arbitrarily reduced the location of the iceline as obtained by our $\alpha$ disk model (Paper I) by 1 AU lead to a inward shift of the ``outer group'' by approximately also 1 AU, bringing it actually to better agreement with the observational data. 

Panel (B) shows the cumulative distribution functions corresponding to (A). The KS test leads to a significance of $\sa=63.9\%$ using the bootstrap samples,  eqs. \ref{eq:significanceKSfromdKS1} to \ref{eq:significanceKSfromdKS3} to 63.5 \%. This confirms the visual impression that the observed and the synthetic semimajor axis distributions differ more than the mass distributions, even if they are still statistically similar. 

When comparing the semimajor axis distribution with the observations, one should keep in mind that this distribution is very likely more affected by long timescale planet-planet interactions (e.g. Ford \& Chiang \cite{fordchiang2007}; Veras \& Armitage \cite{VA04}) than the mass distribution because scattering and ejections are more likely than collisions (Ford \& Chiang \cite{fordchiang2007}), so that it is not necessarily expected that our model can accurately reproduce the observations.  It would therefore be interesting to couple the outcome of a population synthesis calculations to such numerical N-body scattering experiments, similar to the work of Thommes et al. (\cite{thommesetal2008}).
 
 \subsection{Metallicity [Fe/H]}\label{subsect:metallicity}
The fourth distribution we have statistically compared is the metallicity distribution. The so called ``metallicity effect'', \textit{i.e.} the increase of the detection probability with stellar metallicity is discussed in \S\ref{subsect:metallicityeffect}.

It is clear that such a test only makes sense if one can assume that the planets in the observational comparison sample originate from a search within a sample with a similar metallicity distribution as the CORALIE sample. This is  approximately the case for the 1040 FGK stars in the Keck, Lick and AAT planet search sample (FV05, Paper I). It is clearly not the case for detections coming from metallicity biased search programs. This is of course also the case for all other distribution we have compared statistically. As most planets in our observational comparison sample have rather long periods (and are discoveries dating several years back), we assume that our statistical comparison sample is in that sense not ``contaminated''  by planets of a metallicity biased search program (see Santos et al. \cite{santosetal2005} for a discussion of this point).
 
\begin{figure*}
	\centering
      \includegraphics[width=17cm]{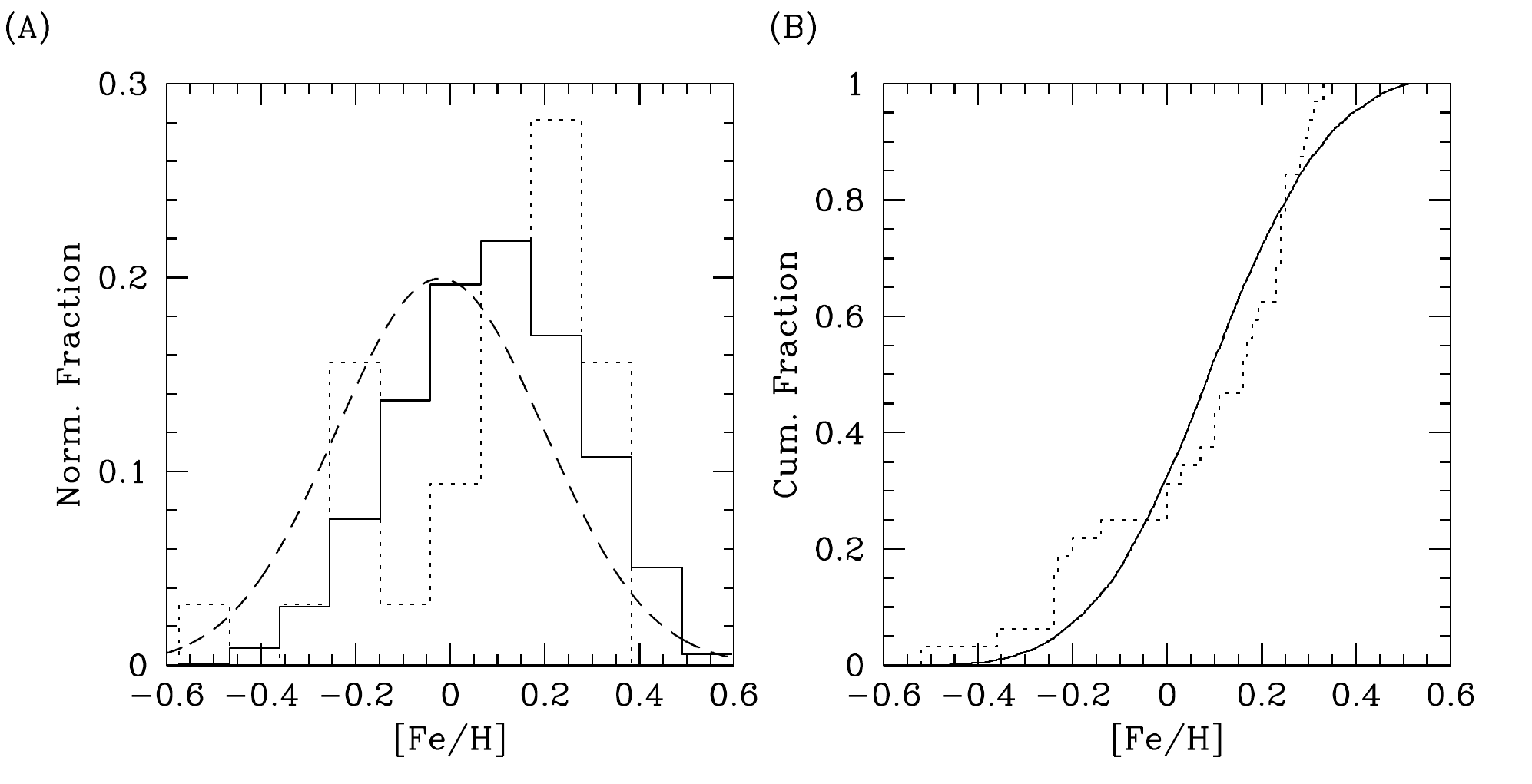}
      \caption{Statistical test of the metallicity distribution. 
      \textbf{Panel (A)}: Solid line: [Fe/H] histogram of the detectable synthetic planets. Dotted line: Distribution in the observational comparison sample. Dashed line: Fit to the metallicity distribution in the CORALIE planet search sample, from which the  $\ninit$ initial conditions are drawn. The distributions of both the synthetic and actual detectable planets are shifted towards higher metallicities relative to this curve.  \textbf{Panel (B)}: Corresponding cumulative distribution functions of the synthetic and actual planets. 
      }  \label{fig:fehist}
\end{figure*} 

Panel (A) of fig. \ref{fig:fehist} displays three distributions: The metallicity distribution of the synthetic detectable planets, the metallicity distribution of the observational comparison sample, and the fit to the metallicity distribution of the CORALIE planet search sample (Santos et al. \cite{santosetal2003}). This distribution is a Gaussian with $\mu=-0.02$ and $\sigma=0.22$ and is the distribution from which we drew [Fe/H] (Paper I), and thus represents the metallicity distribution of all the $\ninit$ initial conditions.

It is an observationally very well established fact that planet host stars have a clearly higher mean metallicity than the complete search sample. The offset between the two distributions is observationally found to be 0.1-0.25 dex (Santos et al. \cite{santosetal2004a}; FV05). Panel (A) shows that our simulations reproduce this observational constraint. Here too, the detectable sub-population is richer in metals than the full population. We find that the detectable sub-populations has a mean [Fe/H] of 0.09 \textit{i.e.} we find a relative shift of 0.11 dex. 

The corresponding cumulative distributions in panel (B) show that the distributions of the actual and synthetic detectable planets are similar. The $\nobsreal=32$ observational comparison sample might be affected by small number effects, as indicated by the bumpy structure of the histogram. It also seems that our synthetic distribution suffers from a certain deficit of detectable planets in the clearly subsolar metallicity regime, while we find more planets at around [Fe/H]=0, where a decrease in the observational data exists. This decrease is however very likely only a small number effect, as indicated by the comparison with a less specific set of exoplanets (\S \ref{subsection:observationalcomparisonsample}). The KS test returns a clearly non-zero, but rather low significance for both samples to come from the same parent distribution of 21.7 \% using the bootstrap method. Using eqs. \ref{eq:significanceKSfromdKS1} to \ref{eq:significanceKSfromdKS3} leads to 23.5\%.

We plan to include in future a self-consistent calculation of the early solid surface density evolution, similar to the model of R{\'o}{\.z}yczka et al. (\cite{rozyczka}) as these authors have shown that this is important for reproducing the observed metallicity distribution of the extrasolar planets.  

\subsection{Observational constraints for the ``Hot'' planets}\label{subsect:obsconsthotjup}
As explained in Paper I, due to possible effects like Roche lobe overflow to the star (Trilling et al. \cite{trillingetal1998}), evaporation or partial accretion of gas streaming past the planet onto the star  we do not really know the fate of the ``Hot'' synthetic planets migrating  to $\atouch\approx0.1$ AU.

Up to this point, ``Hot'' planets were therefore excluded from the statistical analysis. However, Hot Jupiters represent an important feature of the actual population of extrasolar planets, and therefore we still would like to use them to test our models. To be able to do this, we proceeded in the following way: FV05 have published a list of 850 FGK stars  that have enough RV observations that every planet with a radial velocity semi-amplitude $K>$30 m/s and a period $T$ shorter than 4 yr was detected (uniform detectability criterion).  Of the  850 stars, 47 have a planet (or several planets) detectable with this observational bias. Using the planetary orbital parameter from FV05,  we have then calculated how many of the 47 stars with planets have a companion (using the most massive one if there is more than one planet) that would have been classified as a ``Hot'' planet in our simulations,  using the $\atouch$ criterion. This is the case for 9 stars.  We can use this to define a simple new constraint in the following way. With a $K>$30 m/s, $T<4$ yr pass-fail bias, the overall detection probability excluding the ``Hot'' planets $\pfv$ is $38/850\sim 4.5 \%$.  Including  the ``Hot'' planets  leads to a $\pfvwhot=47/850 \sim 5.5 \%$.  The fraction of ``Hot'' detectable planets of all detectable planets is $\ffv=9/47\sim 19 \%$.  This last figure is clearly the most important observational constraint,  as it is a ratio between detectable planets only, which is likely to reduce possible unwanted consequences of the one-embryo-per-disk approach.  We also note that the fraction of all stars with a detectable ``Hot'' giant planet is $9/850\sim 1$\% as in Marcy et al. (\cite{marcyetal2005}) who find $1.3\pm0.3\%$. 

We then proceeded in the same way for the synthetic planets. We sorted out the subset of synthetic planets which have $K>$30 m/s and $T<4$ yr (``FV05 bias'').   The radial velocity semi-amplitude $K$ and $T$ are calculated as (Udry \cite{udry2000})
\begin{eqnarray}\label{eq:velocityamplitude} 
K&=&\frac{\msini}{\sqrt{1-e^2}}\sqrt{\frac{G}{a \mstar}}\\
   &\approx&0.09\left(\frac{\msini}{1 \mearth}\right)\left(\frac{a}{1\mathrm{AU}}\right)^{-0.5}\,\,\mathrm{\frac{m}{s}}\\
T&=& \sqrt{\frac{4 \pi^2 a^3}{G \mstar}}\\
  &=& \left(\frac{a}{1\mathrm{AU} }\right)^{3/2}\,\,\mathrm{yr}
\end{eqnarray} 
where the second lines apply for our special case of a circular orbit and $\mstar= 1 \msun$. 

Such a pass-fail detection criterion is much more rudimentary than the synthetic observational detection probability presented in \S\ref{subsection:syntheticRVbias}, but we want to use a detection criterion as similar as possible to FV05.  For the ``Hot'' planets,  whose real final semimajor axis is unknown we once calculated $K$ assuming that their final semimajor axis is 0.1 AU, and once assuming 0.01 AU, but always using the mass when they arrive at the feeding limit at $\atouch$. The smaller assumed final semimajor axis is of course extremely close to the star (the radius of the sun is about 0.005 AU). But in this way, we obtain an approximative upper and lower boundary.

\begin{table*}
\begin{center}
\caption{Constraints on the "Hot" planets from Fischer \& Valenti (\cite{fischervalenti2005}) and results from the synthetic population. The two columns 0.1 and 0.01 AU represent the results obtained if such final semimajor axes for the synthetic ``Hot'' planets are assumed.}\label{tab:hotjupresults} 
\begin{tabular}{llll}\hline
Feature & 0.1 AU & 0.01 AU & obs. const. (FV05) \\ \hline
Nb. of detectable synth. planets with the FV05 bias outside feeding limit & \multicolumn{2}{c}{3382}  & - \\
Detection probability with FV05 bias excluding ``Hot'' planets ($\pfv$) [\%] & \multicolumn{2}{c}{4.8} & $4.5\pm0.7$ \\
Nb. of ``Hot'' synth. planets detectable with the FV05 bias   &  259 & 407 & - \\
Detection probability with FV05 bias including ``Hot'' planets ($\pfvwhot$) [\%]  & 5.2 &5.4 & $\geq5.5\pm0.8$\\
Fraction of ''Hot'' detectable planets of all FV05 detectable planets ($\ffv$) [\%]&  7.1 & 10.7& $\geq 19 \pm 6 $\\ 
\hline
\end{tabular}
\end{center}
\end{table*}

The results are as follows (tab. \ref{tab:hotjupresults}): Outside the feeding limit, 3382 synthetic planets are detected with the FV05 bias.  This results in  a $\pfv$ of $3382/70\,000\sim 4.8 \%$ for the synthetic population, compatible with the observed value of $4.5\pm0.7\%$, assuming for the error that the observations follow Poisson statistics (as FV05).  Among the ``Hot'' planets, 259 or 407 planets are detected with the  FV05 bias, for $a=0.1$ or 0.01 AU, respectively. This gives us a  synthetic $\pfvwhot$ of (259+3383)/70\,000 to (407+3383)/70\,000 corresponding to 5.2 to 5.4 \%, again compatible with the observations ($5.5\pm0.8$\%). For the most important constraint,  the fraction of detectable ``Hot'' planets among all planets  detectable with the FV05 bias, we find for the synthetic population  $\ffv=$7.1 - 10.7 \% for an assumed final semimajor axis of 0.1 and 0.01 AU, respectively. While the value using 0.1 AU is too low, the upper limit is not too far from the observed ratio of $\geq 19 \pm 6 $\% given the large error bars.  We note that this result corresponds to an overall frequency of synthetic Hot Jupiters of 259/70\,000=0.37 \% and 407/70\,000=0.58 \% again for the lower and the upper limit, respectively.

We note that in contrast to our result Ida \& Lin (\cite{idalin2004a}) have found that a much higher number of planets must have migrated close to the star compared to what is observed today which would correspond to a significantly higher $\ffv$. Such a higher $\ffv$ is also in agreement with the observational constraint, as planets may have perished by falling into the star or by evaporation. However, we have found  in our calculations that we cannot obtain synthetic populations with a higher $\ffv$ (as obtained when assuming a higher type I migration rate, \S \ref{subsubsect:f1variation}) without simultaneously increasing also the number of detectable planets at $a\lesssim1$ AU but still outside the feeding limit. Since the nominal model tends to already overestimate the number of planets in this region, we conclude that it is not possible to increase $\ffv$ without degrading significantly the KS test result for the semimajor axis. This is an example that illustrates clearly that the model cannot be tweaked in a particular direction by a suitable choice of parameters. For the moment, we can only speculate that these difficulties are due to an incomplete description of migration, and/or that our simulations lack a mechanism that ``produces'' Hot  Jupiters without changing the semimajor axis distribution at larger distances. Note that a simple stopping mechanism is not a solution since we actually have kept all planets migrating to these short distances. A mechanism that is controlled by the thermodynamic structure of the disk that could have such an effect is presented in \S\ref{subsubsect:rocklineeffect}.  Alternatively, one could also imagine that the very strong instrumental (and very likely also strong observer's psychological) bias to find Hot Jupiters is not completely corrected for in the observational comparison data.  

Indeed, other surveys have found a rather lower rate of occurrence of Hot Jupiters (less than 1 \%), which we reproduce better with our simulations (0.4-0.6 \%): Naef et al. (\cite{naefetal2005}) find for the ELODIE planet search campaign that a fraction of $0.7\pm0.5\%$ of stars have a giant planet with a period less than 5 days and a mass larger than 0.42 $\mj$. Cumming et al. (\cite{cummingetal2008}) find a similar result of $0.65\pm0.4\%$. Fressin et al. (\cite{fressinetal2007}) find that a fraction of 1/215 = 0.47 \% of late main-sequence stars are orbited by a giant planet with a period between 1-5 days, using combined results from radial-velocity and photometric transit surveys. 

\subsection{Metallicity effect}\label{subsect:metallicityeffect}
\begin{figure}
   \resizebox{\hsize}{!}{\includegraphics{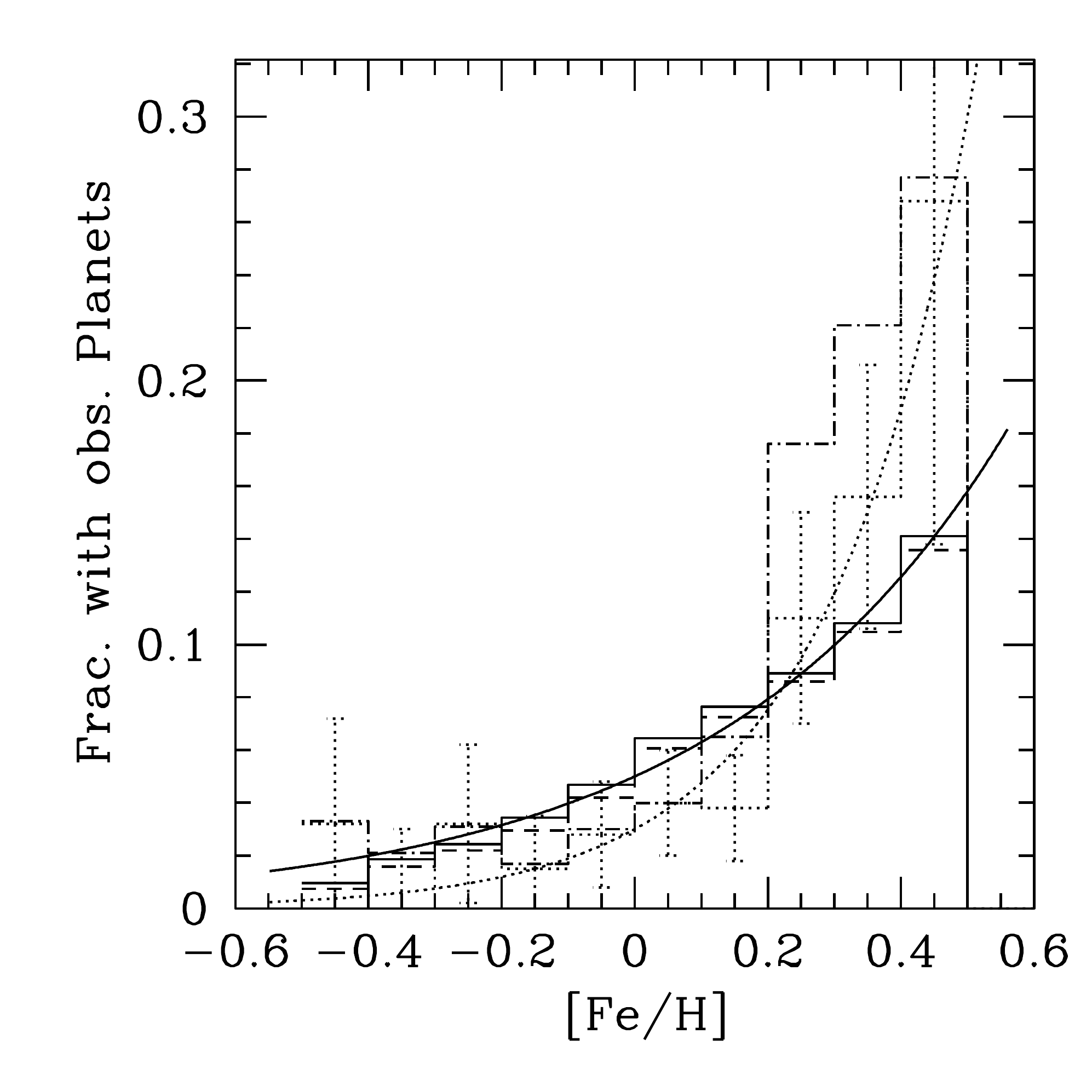}}
   \caption{Fraction of embryos that become giant planets detectable with $K>30$ m/s and $T<$ 4 yrs as a function of metallicity for the synthetic and actual surveys. The solid line assumes that the ``Hot'' synthetic planets stop at 0.01 AU, the dashed one that they do so at 0.1 AU. The dotted lines are the observationally determined values and error bars taken from Fischer \& Valenti (\cite{fischervalenti2005}). The dashed-dotted line is the averaged result from the CORALIE and Lick-Keck-AAT samples (Udry \& Santos \cite{udrysantos2007}). The smooth solid curve is given as $0.05\times10^{[Fe/H]}$ \textit{i.e.} $\propto Z$, which approximately fits our results, while the smooth dotted curve is the $0.03\times10^{2.0\times [Fe/H]}$ \textit{i.e.} $\propto Z^2$ fit from Marcy et al. (\cite{marcyetal2005}). }  
   \label{fig:PdetecfehFV}
\end{figure}

The strong correlation between stellar metallicity, and the likelihood to find giant planets, the so called ``metallicity effect'', is possibly the best established observational constraint (e.g. Gonzales \cite{gonzalez1997}; Santos et al. \cite{santosetal2004a}; Fischer \& Valenti \cite{fischervalenti2005}). Even though the role of metallicity in planet formation will be discussed in detail in a forthcoming publication, we check here if our synthetic population presents the same effect as has been done  for other giant planet formation models (Ida \& Lin \cite{idalin2004b}; Kornet et al. \cite{kornetetal2005}; Robinson et al. \cite{robinsonetal2006}; Matsuo et al. \cite{matsuoetal2007}). In \S \ref{subsect:metallicity} we have already compared the  [Fe/H] distribution of the synthetic and the actual planets. But as explained there, these distributions depend on the metallicity distribution of the planet search sample. By dividing the [Fe/H] distribution of the planet host stars by the one of the full sample we can correct for that. 

To do so, we use the sub-population of  3641 respectively 3789 planets detectable with the FV05 bias described in the last section and assigned them to the same bins in metallicity 0.1 dex wide as FV05. We then normalize by dividing the number in each bin by the total number of initial conditions in the corresponding bin. With this procedure, both synthetic and observed data have been binned in identical manner which allows for an accurate comparison.

Fig. \ref {fig:PdetecfehFV} shows the results of this comparison. It can be seen that the synthetic population reproduces the observed positive correlation between metallicity and the detection probability of giant planets. Even quantitatively it agrees with the observational constraint from FV05 within the error bars in all bins except two at 0.1-0.2 dex.  Note that we have not additionally normalized the distribution at any bins. The fact that we also reproduce the absolute numbers well means that our single embryo is obviously a good representative for the large number of all nascent embryos.

Fig. \ref{fig:PdetecfehFV} makes however also clear that the detection probability as a function of  [Fe/H] in the model and the observational data are not identical. At largely supersolar metallicities, our predicted fractions are below the observed values (but still within the Poisson error given by FV05) while at mildly supersolar values we find slightly higher fractions than FV05. We find that the increase of the detection probability scales approximately as $\propto Z$ ($Z$ being the stellar mass fraction of heavy elements), while observational data rather indicate a quadratic dependence on $Z$ (Marcy et al. \cite{marcyetal2005}), or two different regimes for metallicities larger respectively smaller than 0.0 dex, with no dependence on [Fe/H] in the subsolar regime (Santos et al. \cite{santosetal2004a}; Udry \& Santos \cite{udrysantos2007}). Our theoretical predictions seem to follow approximately the same dependence on [Fe/H] over the entire domain, which is the also the case for Ida \& Lin (\cite{idalin2004b}, \cite{idalin2008b}).

The reasons for this difference in slope could be metallicity effects that are not included in the model. At the moment, the only quantity where $\fpg$ enters is the solid surface density. Metallicity effects that are not included are for example changes in disk and envelope opacity, or a change in relevant planetesimal size and consequent radial distribution. This last effect was found  by  Kornet et al. (\cite{kornetetal2005}) to increase the detectable fraction especially at low [Fe/H].

In any case, it is apparent that the synthetic population does quite well when it comes to reproducing the ``metallicity effect''. The fact that this happens \textit{concurrently} with all other observational constraints, is the critical point in this study. We thus confirm the result of parameterized core accretion models (Matsuo et al. \cite{matsuoetal2007}) that the core accretion paradigm seems to be capable of explaining the vast majority of known extrasolar giant planets. 

A closer look at the population reveals that the metallicity effect is due to the combination of two effects: First, metal rich systems favor the formation of massive planets, as one would intuitively expect from the core accretion theory, as for metal rich disks the core reaches the critical mass for runaway gas accretion earlier, therefore allowing more gas to be accreted before the disappearance of the disk. The second is simply the fact that the RV technique discovers massive planets more easily (fig. \ref{fig:detecprobRV}). It does however not  imply an absence of other types of planets (not Jovian) at low metallicity. 

With the metallicity effect, we have treated all six observational constraints mentioned in \S\ref{subsect:definingobsconstarints}. In the next section, we address an observational constraint that is not posed by the extrasolar planet population itself, but by the protoplanetary disk they are formed in.

\subsection{Formation timescales}\label{subsect:formation timescales}
\begin{figure}
   \resizebox{\hsize}{!}{\includegraphics{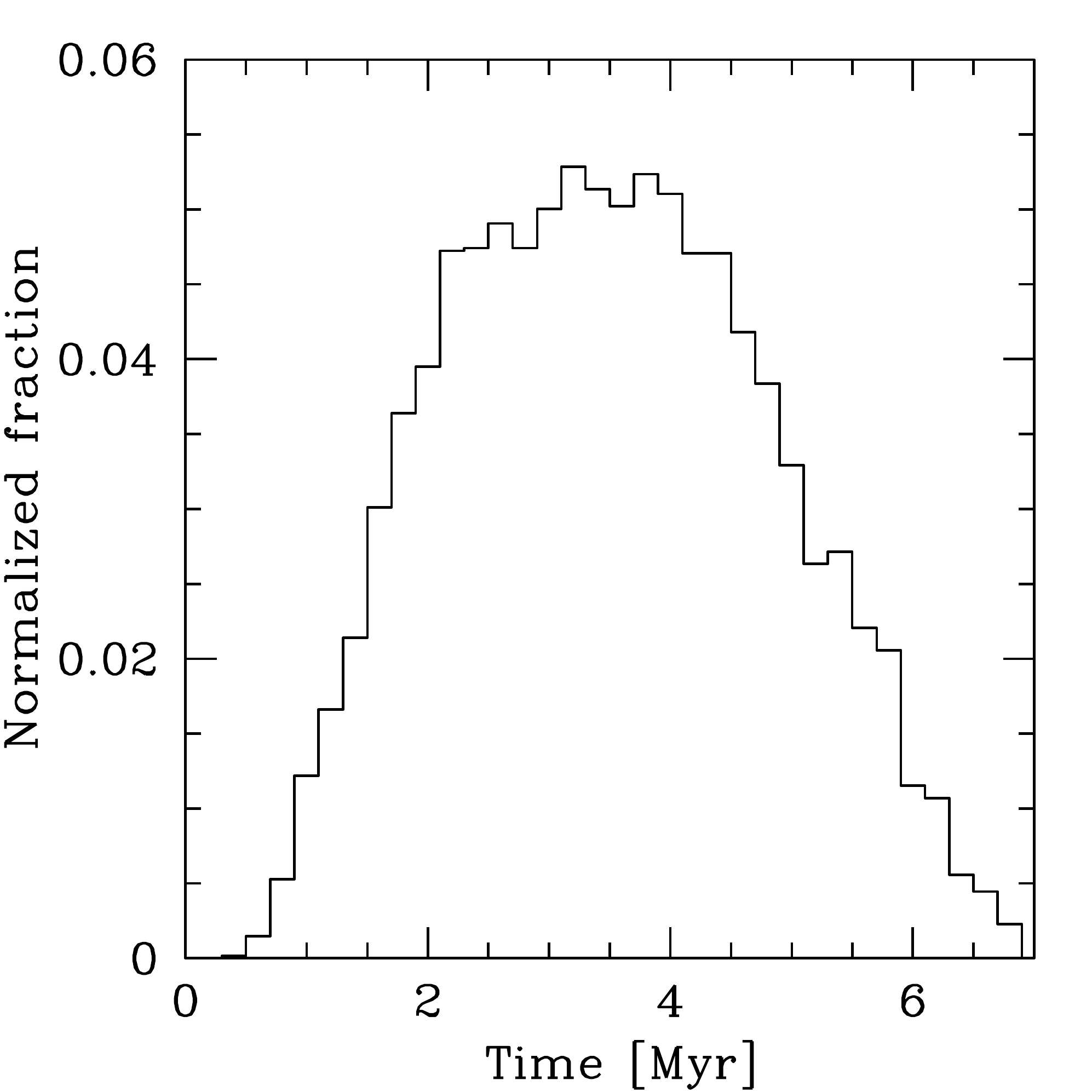}}
   \caption{Distribution of the disk lifetimes in which detectable synthetic planets are formed, which are upper limits for the formation timescales of the detectable (giant) synthetic planets themselves. Detectable planets originate from disks that are in the mean somewhat longer living than average disks, but giant planets can also form in disks that have a lifetime as short as 1-2 Myr.} 
   \label{fig:Tform}
\end{figure}

Early core accretion models (e.g. Pollack et al. \cite{pollacketal1996}) suffered from the so called timescale problem. The predicted formation timescales for giant planets were comparable to, or even longer than typically observed disk dispersion timescales. This lead to the hypothesis that a faster formation mechanism is needed to form Jovian planets, namely the gravitational instability model  (Boss \cite{boss2001}; Mayer et al. \cite{mayeretal2004}).

Later  extensions and improvements of the core accretion model, such as the inclusion of planetary migration and concurrent disk evolution (Alibert et al. \cite{alibertetal2004}, \cite{alibertetal2005a}) or the use of modified grain opacities  in the envelope (Podolak \cite{podolak2003}; Hubickj et al. \cite{hubickyjetal2005}),  showed that the core accretion mechanism is able to form giant gaseous planets well within the timescale limit imposed by the observations. On the other hand, Fortier et al. (\cite{fortieretal2007}) have recently shown that when they employ a core accretion rate based on the calculations by Thommes et al. (\cite{thommesetal2003}) for the oligarchic growth regime, they find for identical initial conditions a formation time for Jupiter about one order of magnitude larger than Pollack et al. (\cite{pollacketal1996}), so that this question is not yet settled (see Paper I).  In the population synthesis calculations presented here which use the same (faster) accretion rates of solids as Pollack et al. (\cite{pollacketal1996}), the formation timescales of giant planets are by construction compatible with the observed disk lifetimes, as the distribution of the photoevaporation rate $\mwind$ was adjusted to reproduce the observed disk lifetime distribution (Paper I), and disk lifetimes are obviously upper limits for the formation timescales of gaseous planets (as we neglect any formation after the gas disk dispersion, the disk lifetimes are in fact an upper limit for the formation timescales of all planets in the model).

A posteriori, the Monte Carlo simulations yield the distribution of lifetimes of those disks which eventually produced a giant planet in the detectable sub-population. Fig. \ref{fig:Tform} shows this distribution.  We find that giant planets form in disks with a mean lifetime that is about one Myr longer than the mean lifetimes of the disks of all $\nsynt$ planets (mean lifetimes of 3.50 and 2.49 Myr, respectively).  Some giant planets ($\sim14.5$\%) were however also formed in short-lived disks ($<2$ Myr). This provides a hint that from the disk lifetimes aspect alone, giant planet formation is not be completely inhibited (albeit less likely) in a dense stellar environment in which circumstellar disk dispersion is rapid. However, for this to occur, high amounts of solids are required: The mean metallicity of the disks in which giant planets form within 2 Myr is 0.18 dex, twice as large as the mean metallicity of all detectable giant planets (0.09 dex, as mentioned in \S \ref{subsect:metallicity}).

\subsection{Influence of parameters}\label{subsect:influeceofparameters}
The results presented up to this point relate to the nominal population. This nominal population was defined as the population that provided an overall best match to all the tests described above from a large set of populations that were obtained from varying  parameters.  Table \ref{tab:variedparameters} summarizes the values of some parameters and distributions that were used to explore all space available. The synthetic populations obtained with these different parameters were compared to the actual exoplanets using the same statistical methods as  discussed above.
\begin{table}
\begin{center}
\caption{Important parameters for the simulations. If several were tested, the value for the nominal case which gives the best results in the statistical analysis is printed in italic letters.}\label
{tab:variedparameters}
\begin{tabular}{ll}\hline
Feature & Values\\\hline
Type I migration efficiency factor $\f1$  & \textit{0.001}, 0.01, 0.1, 1 \\
Viscosity parameter $\alpha$ &  0.001,  0.005, \textit{0.007}, 0.01  \\ 
Initial gas disk profile $\Sigma(a,t=0)$ & $\propto a^{-3/2}$\\
Rockline included & yes, \textit{no}\\
Iceline included & yes\\
Photoevaporation included & \textit{yes}, no \\
$f_{\rm D/G,\odot}$ & 0.016, 0.02 \textit{0.04}, 0.05\\
Distribution for $\sigmanorm$ & Taurus, \textit{Ophiuchus} \\
Host star mass & 1 $\msun$\\
Initial embryo mass $\membstart$ &0.1, 0.3, \textit{0.6} $\mearth$\\
\hline
\end{tabular}
\end{center}
\end{table}
In this section, we  briefly discuss some outcomes of varying various free parameters. Our main focus rests on determining the influence of changing the  type I migration efficiency factor $\f1$ since this has the largest influence on the  sub-population of ``Hot'' planets.

\subsubsection{Varying $\sigmanorm$ and $\membstart$}\label{subsubsect:varyingmwindsigmaetc}
In all cases mentioned here, only one parameter was changed from its default value at one time.

Using the disk mass distribution derived from observational data of the Taurus-Auriga star forming region (with $\mu=-1.66, \sigma=0.74$, cf. Paper I) instead of Ophiuchus for $\sigmanorm$ does not have a marked influence on our results. In the KS tests, values of 65, 85, 58 and 50\% are found for $\sam, \sm, \sa$ and $\sfeh$, respectively.  The reason for these rather small changes is partially linked to the fact that we kept the range of possible $\sigmanorm$  values constant (50-1000 g/cm$^2$), so that in both cases  a rather limited part of the full distribution is covered (Paper I).  

We have also synthesized a population with $\membstart = 0.3$ instead of 0.6 $\mearth$. As the starting mass has been chosen arbitrarily, changing it within certain limits should not have a significant influence on the final results.  A different $\membstart$ changes the distributions of $\astart$ (via the isolation mass criterion) and naturally also $\tstart$. In addition lowering $\membstart$ increases the number of disks that can form such embryos within their lifetime. The fraction of initial conditions $\nnocalc/\ninit$ drops from 28\% to 21\%. Qualitatively, the differences to the nominal case are very small concerning the structure of the mass-distance diagram. As for the KS tests, the largest difference is obtained in the mass distribution ($\sm=47$\% instead of 95.6\%). At first glance, this seems to be a large change and shows that the KS tests are quite sensitive and large differences can occur while ``by eye`` changes appear quite small. In many other cases however, the KS significance was found drop much more e.g. to $\lesssim0.1\%$. The other KS results are 53, 57 and 37\% for $\sam, \sa$ and $\sfeh$, respectively.

\subsubsection{Varying $\f1$}\label{subsubsect:f1variation}
\begin{figure*}
\centering 
\includegraphics[width=17cm]{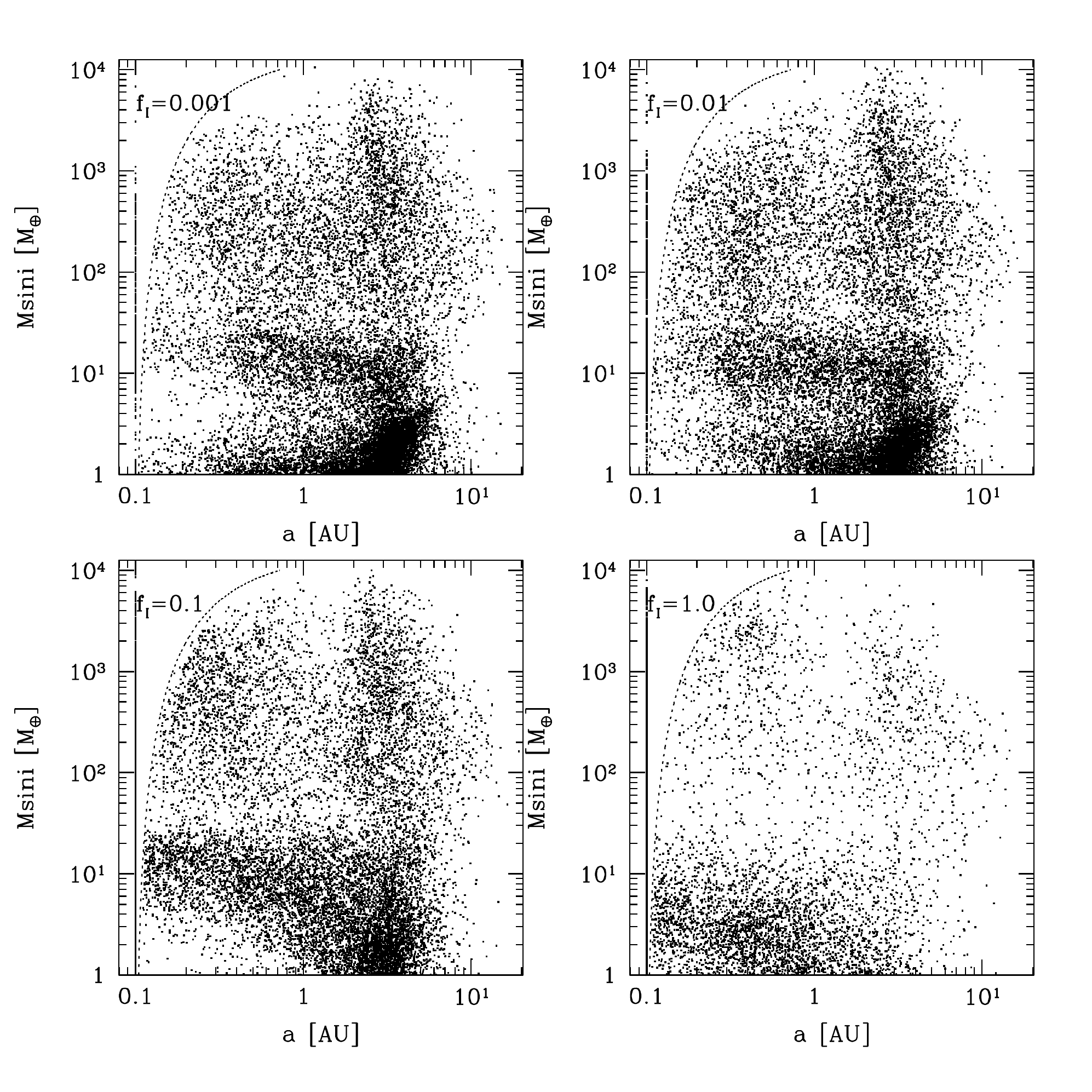}
\caption{Semimajor axis $a$ versus projected mass $\msini$ for four synthetic populations differing in the type I migration rate efficiency factor $\f1$. In all cases, 20\,000 synthetic planets are plotted. The dotted line is the feeding limit at $\atouch$. Planets migrating to this limit have been plotted at 0.1 AU. The $\f1=0.001$ case (top left), is the nominal case \textit{i.e.} the plotted planets are simply a subset of those in fig. \ref{fig:amdist}, panel A.} 
\label{fig:f1evopathes_compo}
\end{figure*}

The current knowledge about type I migration which embedded low mass planets undergo is rather shaky. For example, Nelson \& Papaloizou (\cite{NelsonPap04}) have shown that migration of low mass planets in a  disc with magnetohydrodynamic turbulence occurs in the form of a random walk, and that  averaged torques don't converge to well-defined values for low mass planets.  Menou \& Goodman (\cite{menou}) have found a strong sensitivity of local migration rates to the background disk structure like opacity transitions. Paardekooper \& Mellema (\cite{paardekoopermellema2006}) and Baruteau \& Masset (\cite{baruteaumasset}) have also shown that depending on specific local thermodynamic disk properties, the net torque can also be positive, \textit{i.e.} causing outward migration. In addition, poorly ionized ``dead zones'' in the gaseous disk with a very low effective viscosity may also  play an important role (Ida \& Lin \cite{idalin2008b}). For example, they allow much smaller planets to open a gap and to migrate in type II migration characterized by very low $\alpha$ values rather than in type I (Matsumura et al. \cite{matsumuraetal2007}).  

In summary, it seems increasingly likely that type I migration rates found in laminar, isothermal disks (Tanaka et al. \cite{ Tanaka}) might not be the whole story and that the true migration rate of small mass planets might be much more complicated. Nevertheless, short of a better description, we use the rates given by Tanaka's et al. (\cite{Tanaka}) and multiply them by an arbitrary efficiency factor $\f1$. We do not claim that this factor captures the true type I migration rate. It merely provides a convenient way to investigate the changes in the characteristics of the synthetic planet population with changing migration rate (see also Ida \& Lin \cite{idalin2008}). 

We have synthesized three populations which differ from the nominal case with $\f1=0.001$ by the magnitude of the type I efficiency factor $\f1$, for which we use values of 0.01, 0.1, and 1.0. In Paper I, planetary formation tracks for two different $\f1$ cases were shown (0.001 and 0.1), and some effects on the the planetary populations were discussed. Figure. \ref{fig:f1evopathes_compo} shows the final positions of 20\,000 planets in the $a-\msini$ plain for the population with $\f1$=0.001, 0.01, 0.1 and 1.0.  

When generating initial conditions, we don't include planetary migration, as explained in Paper I. This implicitly means that we neglect the radial drift of the embryos at masses below $\membstart=0.6$ $\mearth$. For low values of the efficiency factor $\f1$ (0.001-0.1), this is justified, as the type I migration rate increases linearly with planet mass (e.g. Alibert et al. \cite{alibertetal2005a}), so that the extend of migration a seed would undergo while its mass is below $\membstart$ is only small. This can be seen in the formation tracks in Paper I.  For $\f1=1.0$ however this approximation is no longer correct, as illustrated by the results of Ida \& Lin (\cite{idalin2008}). In order to attenuate the artificial reduction of type I migration we have therefore synthesized the $\f1=1$ population with  an initial mass of the seed of 0.1 instead of 0.6 $\mearth$. 

At $\f1=0.01$ (top right panel), the synthetic population is in general similar as in the 0.001 case. A separation between the ``main clump'' and the ``outer group'' has however become visible, and the close-in very low mass planets ($a\lesssim1$ AU, $m\lesssim 4$ $\mearth$), for which the effects of type I migration are the most severe (see also Ida \& Lin \cite{idalin2008}) have started to migrate and grow beyond the isolation mass, and sometimes also have reached the feeding limit. 

At $\f1=0.1$ (bottom left panel) the effects of the type I migration start to become clearly visible compared to the nominal case. The sub-population of ``failed cores'' is clearly reduced at large distances. Most embryos starting inside the iceline have fallen into the feeding limit,  and only ``failed cores'' starting late and beyond the iceline are retained. Nevertheless, some features of the nominal populations like the ``outer group'' and the ``horizontal branch'' are visible also at this higher migration rate (by a factor 100). The inner boundary of the ``main clump'' is shifted inwards and reaches to the inner boarder of the computational domain.  A significant number of planets with masses up to 20 $\mj$ have now migrated to the feeding limit. The presence of these massive planets is one of the reasons for which we rejected high values of $\f1$ since these objects should have been easily detected. In fact, such planets have only been detected orbiting host star which are members of multiple stellar systems, but not around single host stars (Udry et al. \cite{udryetal2003}; Eggenberger et al. \cite{eggenbergeretal2004}).

The faster type I migration has also, as expected, consequences for the low mass planets reaching the feeding limit. In particular, with increasing value $\f1$ smaller and smaller mass planets migrate and eventually reach 0.1 AU, For example, namely 6 and 2 $\mearth$ planets reach 0.1 AU for $\f1=0.1$ and 1.0, respectively. In the case of the nominal population only planets with a mass larger than about 10 $\mearth$ reach the inner boundary of the computational disk. In addition, the nominal model shows a depletion of planets in the region between the ``failed cores'' which virtually do not migrate, and those in the ``horizontal branch'' as discussed in \S \ref{subsect:aMdiagram}. This depletion disappears at higher values of $\f1$. 

Finally, we note that even for $\f1=1.0$ giant planets can form (Thommes \& Murray \cite{thommesmurray2006}), but only at a reduced number (tab. \ref{tab:f1variationresults}). The sub-population of ``failed cores'' which was very numerous just beyond the iceline at small type I rates has almost completely migrated inside 1-2 AU or even into the feeding limit. This has strongly populated a part of the $a-\msini$ plane ($\msini \sim 4$ $\mearth$, $a \lesssim 0.3$ AU) which is depleted at low $\f1$. 

The presence or absence of essentially gas-free, close-in icy or ``ocean'' planets (L\'eger et al. \cite{legeretal2004}) with masses below 10 $\mearth$ for which we can exclude that they have lost a significant primordial atmosphere could be used as a strong indicator of the efficiency of type I migration.  Indeed we find in the model that planets which are (1) mainly icy ($>50$ \% of the accreted solids are icy), (2) which have a mass smaller than 10 $\mearth$, (3) a $\menv/\mcore$ that is much less than 0.1 and (4) which migrate from beyond the iceline to $a \lesssim 0.1$ AU only exist if $\f1=1$. At $\f1=0.001$ we also find planets in the feeding limit which have a mainly icy core, but these planets have a mass of at least 20 $\mearth$, and have more massive envelopes $0.1\lesssim \menv/\mcore \lesssim 1.0$, \textit{i.e.} they are of a Neptunian nature. Hence, the discovery in large numbers of low mass ``ocean'' planets would indicate that the type I migration as described in our model is not correct because the existence of such planets requires a high efficiency for this migration while the more massive and distant planets call for a sharply reduced rate. 

We have statistically compared all different $\f1$ populations with the observational comparison sample. Table \ref{tab:f1variationresults} gives an overview of the results of the KS tests. 
\begin{table*}
\begin{center}
\caption{Statistical comparison of synthetic populations calculated with different type I migration efficiency factors $\f1$. All other parameters are kept constant, except that for the $\f1=1.0$ the seed mass is 0.1 instead of 0.6 $\mearth$. The quantities  are the same as in table \ref{tab:basicresutls} and \ref{tab:hotjupresults}.  Where possible, observational constraints are also given (Naef et al. \cite{naefetal2005}; FV05). \label{tab:f1variationresults}} 
\begin{tabular}{llllll}\hline
Feature & $\f1=$ 0.001 & 0.01 & 0.1 & 1.0 &Obs. \\\hline
Detection probability w/o $\nhot$ ($P$) [\%] & 8.7 & 9.4 & 8.8 & 3.1 & $\sim5-10$\\ 
Detection probability w. $\nhot$ ($\pwhot$) [\%] & 10.7 & 14.0 & 28.7 & 48.0 & $\geq7.3\pm1.5$ \\ 
Fraction of cases migrating to $\atouch$ $(\nhot/\ninit)$ [\%] & 2.0 & 4.7 & 19.9 &  45.0 & - \\
Significance KS  $a-\msini$ ($\sam$)   [\%] & 87.7 & 35.2 & 7.8 & 0.3 & -\\
Significance KS $\msini$ ($\sm$) [\%] &  95.6 & 93.6 & 57.3 & 5.0 &-\\
Significance KS $a$  ($\sa$) [\%] & 63.9 & 18.4 &6.7 & 0.2 &-\\
Significance KS [Fe/H] ($\sfeh$) [\%] & 21.7  & 40.8 & 28.0 & 10.0&-\\ \hline
Detect. prob. with FV05 bias excluding ``Hot'' planets ($\pfv$) [\%] & 4.8 & 5.3 & 5.5 & 2.2&$4.5\pm0.7$\\
Detect. prob. with FV05 bias including ``Hot'' planets ($\pfvwhot$) [\%] &5.2-5.4 &6.0-6.3 & 7.5-8.4 & 4.1-5.5&$\geq5.5\pm0.8$\\
Fract. of ``Hot'' detectable planets of all FV05 detectable planets($\ffv$)  [\%] & 7.1-10.7 &  11.4-16.4 & 26.4-34.8 & 47.1-60.5&$\geq19\pm6$\\ 
\hline
\end{tabular}
\end{center}
\end{table*}

The inspection of tab. \ref{tab:f1variationresults} shows that both the results for $\f1=0.001$ and $\f1=0.01$ are all in at least fair agreement with the observations, with the nominal, slower migration case having somewhat better results for $\sam$ and $\sa$, and the population with $\f1=0.01$ reproducing better the constraints derived for the ``Hot'' planets, and the metallicity distribution, $\sfeh$. At $\f1=0.1$ the mass is still fairly well reproduced, but the significances for $a-\msini$ and $a$ are clearly reduced. At $\f1=1.0$ finally, all KS results have fallen to low or very low values, in particular for the semimajor axis distribution. The main reason for this is, as mentioned, that too many giant planets end up at distances $\lesssim$ 1 AU when the type I migration rate is high. Additionally, the period/semimajor axis gap between the ``main clump'' and the ``outer group'' which opens up at larger $\f1$ is not compatible with the observations either.  The gap opens because the seeds that start inside the iceline in metal rich disks and which later become giant planets, now end up at smaller final semimajor axes than in the slow type I case, where they populate the region around 1 AU. Therefore, the comparison of  the synthetic populations with giant planets found by rather low precision (10 m/s) RV surveys constrain the effective type I migration rate to be $\lesssim1/100$ of Tanaka's et al. (\cite{Tanaka}) result, although also a reduction factor of $\lesssim1/10$ can probably not be completely rejected.

Table \ref{tab:f1variationresults} shows that the detection probability $P$ that does not include the planets that reach the feeding limit is nearly independent of $\f1$ between 0.001 and 0.1, and only decreases by a factor of about 3 for $\f1$=1. It is then down at 3.1 \%, which is a factor 2-3 lower than the observationally determined percentage. For the detection probability using the simpler FV05 bias ($\pfv$) the same behavior is found. This is due to the fact that higher migration rates have two partially compensating effects: While a high type I rate drains seeds into the feeding limit (which reduces the detection probability), it also facilitates their growth because they get quicker into regions where new planetesimals can be accreted (which increases the detection probability). The results here show that for a large span of values for $\f1$, the two effects approximately compensate each other. Only for the population with $\f1=1.0$, the first effect dominates. Note that this prediction is in contrast with Ida \& Lin (\cite{idalin2008}) who have found that the fraction of stars with giant planets is a monotonically decreasing function of the type I migration speed, unless the effect of an enhancement of the solid and gas surface density near the iceline due to a ``dead zone'' is included (Ida \& Lin \cite{idalin2008b}).

The detection probability where we have assumed that all embryos that reach the inner boarder of the computational disk become detectable, regardless of their mass as it would be the case for a ``perfect'' detection technique ($\pwhot$), increases strongly with $\f1$, because the fraction of initial conditions which ultimately lead to a planet migrating to $\atouch$ also increases rapidly with $\f1$: For the nominal population $\nhot/\ninit$ is just 2.0 \%, but at $\f1=1.0$ 45 \% of all seed embryo fall into the feeding limit \textit{i.e.} roughly 20 times more. The fraction of Hot Jupiters detectable with the FV05 bias \textit{i.e.} massive planets in the feeding limit in contrast changes only by a factor $\sim5$ with $\f1$: It is 0.4-0.6, 0.7-1.0, 2.0-2.9, 1.9-3.3, for $\f1$=0.001, 0.01, 0.1 and 1.0, respectively, where the two values again assume a stopping distance of 0.1 or 0.01 AU.  This means that a high type I migration rate increases a lot the number of close-in low mass planets (Hot Super Earth and Hot Neptunes), but not that much the number of close-in giant (Hot Jupiters) planets, as we expect it. To quantify this, we study below the initial mass distribution of the ``Hot'' planets.

We note finally that even if our results favor a low type I migration rate, a non-reduced type I migration rate does not prevent the formation of all giant planets, as mentioned before. This result could appear at odd with some of our former results (e.g. Alibert et al. \cite{alibertetal2005a}), where a reduced type I migration was a pre-requisite for the retention of giant planets. However, in the afore-mentioned calculations, the embryos were introduced in the protoplanetary disk at $\tstart=0$, neglecting their formation time, which led to a phase of rapid migration during the early evolution phases of the disk. In the present models, such an initial phase of type I migration is avoided, since embryos are introduced in the disk after a few Myr, when the disk surface density, and therefore the type I migration rate, have decreased. This is in line with the results of Thommes \& Murray (\cite{thommesmurray2006}).

\subsubsection{The IMF of the ``Hot'' planets}\label{subsubsect:imfhotpla}
From the mass-distance plots for the four type I migration rates, and the considerations made before, one already guesses that the mass histogram of the planets reaching the feeding limit at $\approx$0.1 AU must be significantly different for the four $\f1$ cases.  As the sub-population of planets close to the parent star (including also very low mass objects)  is the one that will become accessible to observations (both high accuracy RV and  transit surveys from space as CoRoT or Kepler) more quickly than planets further away, it is of interest to study the IMF of these planets.  We again express the caveat that when comparing the synthetic mass spectra with the results of upcoming surveys, one must keep in mind the many uncertainties affecting the ``Hot'' planets mentioned in Paper I, in particular those of small mass. For example we neglect subsequent evaporation, and only address \textit{one} (inward migration) of several other possible formation channels for low mass, close-in planets like in-situ formation or shepherding by giant planet migration. See Raymond et al. (\cite{raymondetal2007}) for an overview, and also Kennedy \& Kenyon (\cite{kennedykenyon2008}).  

Note also that here no observational bias has been applied and that the stellar mass is still fixed to $1$ $\msun$. These points will be addressed in a later work. Nevertheless, the mass spectrum of the ``Hot'' planets could be a good tool to better understand type I migration. 

\begin{figure}
	\centering
	 \resizebox{\hsize}{!}{\includegraphics{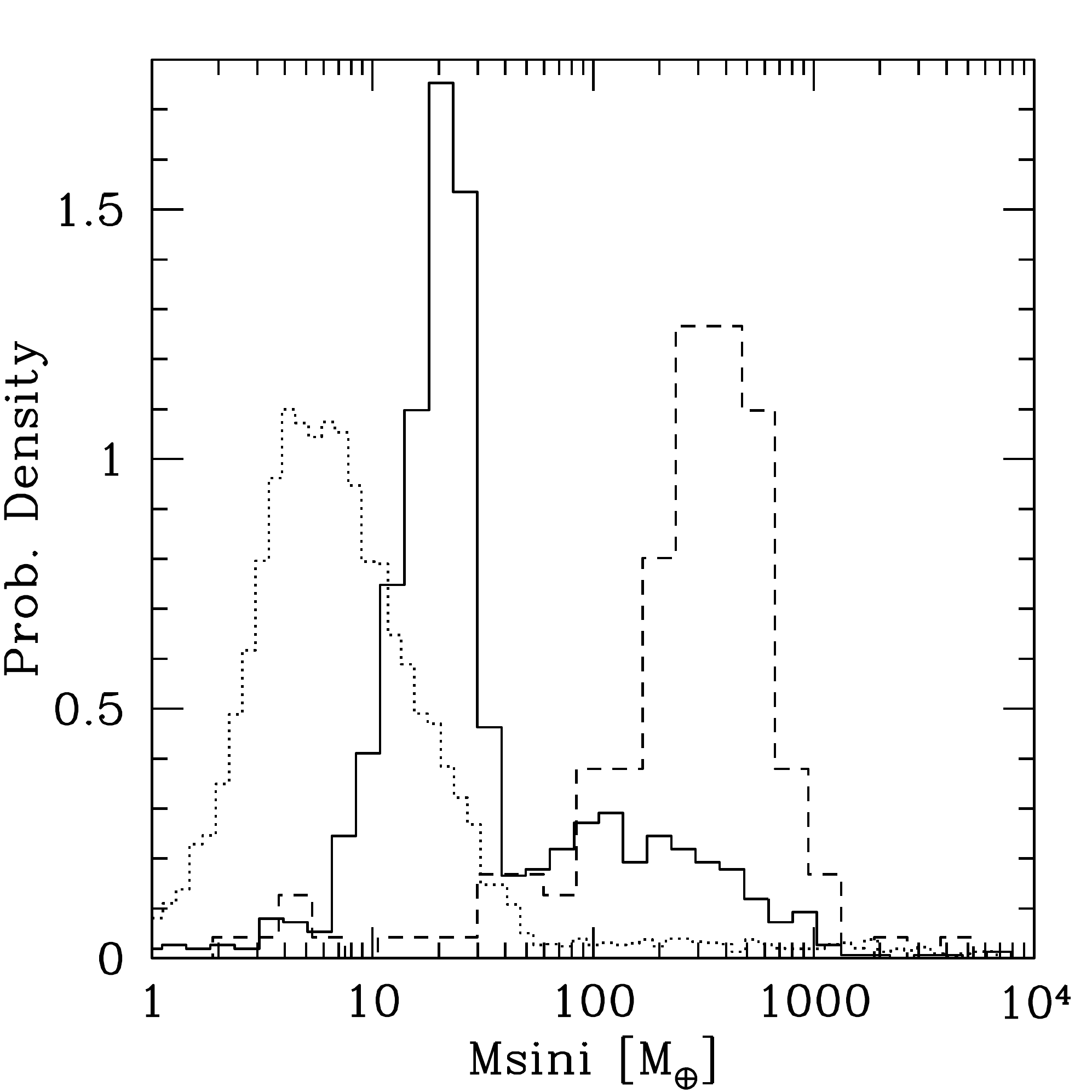}}
      \caption{Distributions of the projected mass of the ``Hot'' planets (\textit{i.e.} those that migrate to the inner boarder of the computational disk), for three different populations:  The solid line is the nominal population with $\f1=0.001$. The dotted line is for $\f1=1.0$.  The dashed line finally is for $\f1$=0.001 too, but with the effect of the rockline included (see \S\ref{subsubsect:rocklineeffect}).  The mass spectra show the result of only one (inward migration) of several other possible formation scenarios for close-in planets like in-situ formation, which are not included in the model. } 
      \label{fig:mhotf1dists}
\end{figure}

In fig. \ref{fig:mhotf1dists} the distributions of $\msini$ of the ``Hot'' planets are plotted for $\f1=0.001$ and 1.0. The other two $\f1$ cases lie between these curves. The third curve in the figure is discussed in \S \ref{subsubsect:rocklineeffect}. In both cases a peak at low masses is seen which corresponds to the Neptunian peak found in the overall (all semimajor axes) IMF (fig. \ref{fig:imf}).  It is the result of planets migrating inwards in the ``horizontal branch''. The exact location and the smallest mass inside the feeding limit is systematically varying with $\f1$:  The higher $\f1$, the more the mass peak is shifted to lower masses and the more it broadens. The higher $\f1$ also, the lower the mass of the lowest mass planet in the ``Hot'' population.  At $\f1=0.001$, the peak of the distribution lies at about 20 $\mearth$, and few planets smaller than $\sim 10$ $\mearth$ have been brought in. For the fast type I migration case, the peak lies at about 5 $\mearth$, and the smallest  planet brought into the feeding limit has a projected mass below 1 $\mearth$. 

Figure \ref{fig:mhotf1dists} also shows that the ratio of massive Hot Jupiters to small Hot Neptune and Hot Super Earth planets changes systematically with $\f1$, as mentioned in the previous section. To study this, we separate the ``Hot'' planets in a high and a low mass bin, using as in table \ref{tab:planettypesbymassfractions} a separating (projected) mass of 30 $\mearth$, as we have identified a local minimum of the overall IMF at approximately this mass\footnote{Also the IMF of the ``Hot'' planets has a local minimum between the planets which do respectively do not undergo gas runaway accretion and is thus bimodal. It lies for the nominal population at a slightly larger mass than in the overall IMF, namely at $M\sim 40$ $\mearth$.}. We then find that there are 2.0, 3.2, 5.1 and 11.4 times more planets in the low mass than in the high mass bin for $\f1$=0.001, 0.01, 0.1 and 1.0. This clear dependence, as well as the dependence of the absolute number of close-in low mass planets on $\f1$ discussed in the previous section will soon be observationally determined.  Recent observational results indicate that close-in low mass planets are very common (Mayor et al. \cite{mayoretal2008}).  

\subsubsection{Heavy element content of ``Hot'' planets}\label{subsubsect:heavyelementhots}
The combined measurements of the mass and radius of planets transiting their host star allow, at least within certain limits, to determine the relative fraction of hydrogen and helium and of heavy elements in their interior, which can for example be used to deduce constraints on the formation of the transiting planet (Figueira et al. \cite{figueiraetal2008}). 

As the number of transiting exoplanets is growing quickly, it is interesting to check if there are statistical correlations between stellar properties and the planetary composition. Indeed, as shown by Guillot et al. (\cite{guillotetal2006}), and confirmed by Burrows et al. (\cite{burrowsetal2007}) and Guillot (\cite{guillot2008}), internal structure models indicate that there is a positive correlation between the total amount of heavy elements in a planet $\mztot$ and the host star metallicity, provided that some `missing physics'  (which can be a modification of the equation of state, a higher opacity or an additional energy source) are assumed to be at work in a similar way in all transiting giant planets.

\begin{figure}
	\centering
	 \resizebox{0.94\hsize}{!}{\includegraphics{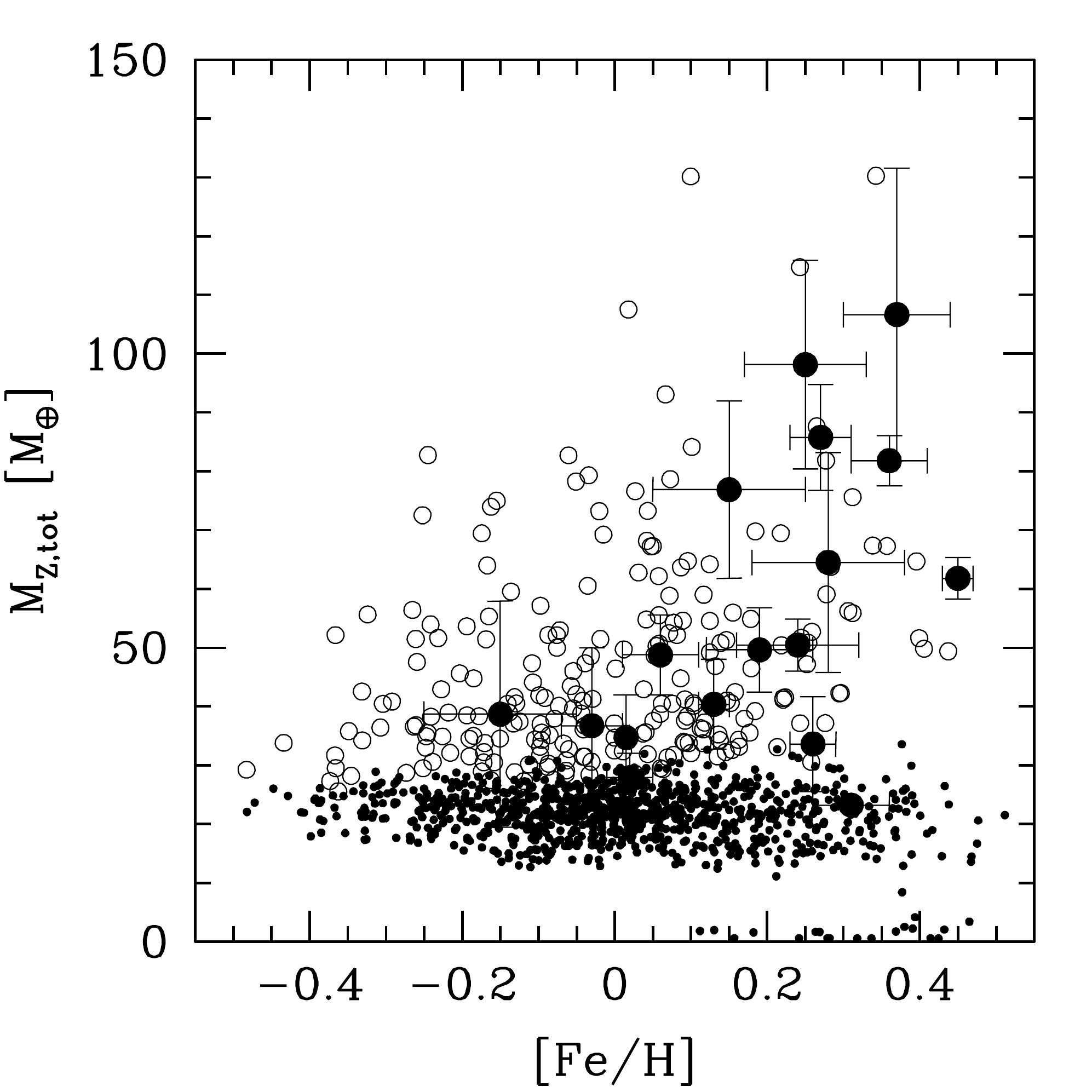}}
      \caption{Total mass of heavy elements of ``Hot'' planets as a function of [Fe/H]. Large empty circles are synthetic planets in the feeding limit with a total mass larger than 100 $\mearth$. Small filled circles are synthetic planets with a total mass below this limit. The large filled circles with error bars are taken from Guillot (\cite{guillot2008}) and show $\mztot$ calculated for 16 currently known transiting exoplanets, assuming a kinetic energy mechanism. HAT-P-11-b was added at [Fe/H]=0.31 and $\mztot\sim23$ $\mearth$ (Bakos et al. \cite{bakosetal2009}).} 
      \label{fig:tristanhotobs}
\end{figure}

Our results for the composition of  the synthetic planets of the nominal population ($\f1=0.001$) reaching the feeding limit are as follows: There is no correlation of [Fe/H] with the maximal total mass (envelope and heavy elements) $M$ or the maximal mass of accreted planetesimals $\mheavy$. But, as shown by fig.  \ref{fig:tristanhotobs}, there is a positive correlation of the total maximal mass of heavy elements $\mztot$ (mass of accreted planetesimals plus mass of heavy elements accreted with the gas).  This is shown by an absence of empty circles representing ``Hot'' planets with $M>100 \mearth$ (as a first order approximation of the detection bias towards massive planets) with a high $\mztot$ at low [Fe/H]. This is similar to the results obtained by Guillot  (\cite{guillot2008}) shown with large black circles for the scenario of an additional energy source deep in the planet's interior. These observed planets also have masses larger than 100 $\mearth$.  A scenario with increased opacities gives qualitatively similar results. The synthetic population also reproduces the general finding of internal structure modeling that some transiting planets contain high amounts ($\gtrsim100 \mearth$) of heavy elements. Note that the host stars in the sample of Guillot  (\cite{guillot2008}) have masses between 0.8 and 1.3 $\msun$, while we have a fixed $\mstar=1\msun$.

The models of Guillot  (\cite{guillot2008}) assume a concentration of the heavy elements in the core surrounded by an envelope with solar composition. Our results of a higher $\mztot$ at high [Fe/H] is however due to higher amounts of heavy elements in the envelope, and not in the core. It could therefore seem questionable if a comparison can be made, as in general the repartition can have a important influence on the radius (Baraffe et al. \cite{baraffechabrierinprep}). The ``Hot'' synthetic giant planets of primary interest here have however a $\mztot/M\lesssim0.3$ (typically $\sim0.1$ as Jupiter and Saturn) so that the influence of the repartition of the heavy elements only has minor influence on the radius predictions (Baraffe et al. \cite{baraffechabrierinprep}).

The small filled circles in fig. \ref{fig:tristanhotobs} represent mainly the abundant sub-population of Hot Neptunes discussed earlier. They typically have $0.7\lesssim\mztot/M\lesssim1$. The only  currently known transiting planet around a solar like star of this type  (HAT-P-11 b, Bakos et al. \cite{bakosetal2009}, $\mztot\sim23\mearth$) is also shown in the figure and falls well in this sub-population where observational biases are still very important.

We have excluded in fig. \ref{fig:tristanhotobs} the handful of extremely high mass planets ($\gtrsim 10 \mj$) in the feeding limit which are in the nominal population quite well separated from the rest of the ``Hot'' giant planets  ($M\lesssim3\mj$). The reason is that due to their extreme mass, their $\atouch$ is very large (several tenths of an AU) so that it is not clear if they can be classified as ``Hot'' planets. In any case, these planets which start their formation inside the iceline in extremely metal rich disks ([Fe/H]$\gtrsim0.4$ and $\sigmanorm\gtrsim 800$ g/cm$^{2}$) are extreme in terms of composition also: They typically have a $\mheavy$ of 100-200 $\mearth$ and $\mztot$ of 200-800 $\mearth$, reminiscent of the internal composition deduced by Baraffe et al. (\cite{baraffechabrierinprep}) for  \object{HD 147506 b} (aka HAT-P-2b, Bakos et al. \cite{bakosetal2007}) where a higher disk mass (due to the higher primary mass) could compensate the lower [Fe/H].

\subsubsection{The ``rockline effect''}\label{subsubsect:rocklineeffect}
\begin{figure}
	\centering
	 \resizebox{\hsize}{!}{\includegraphics{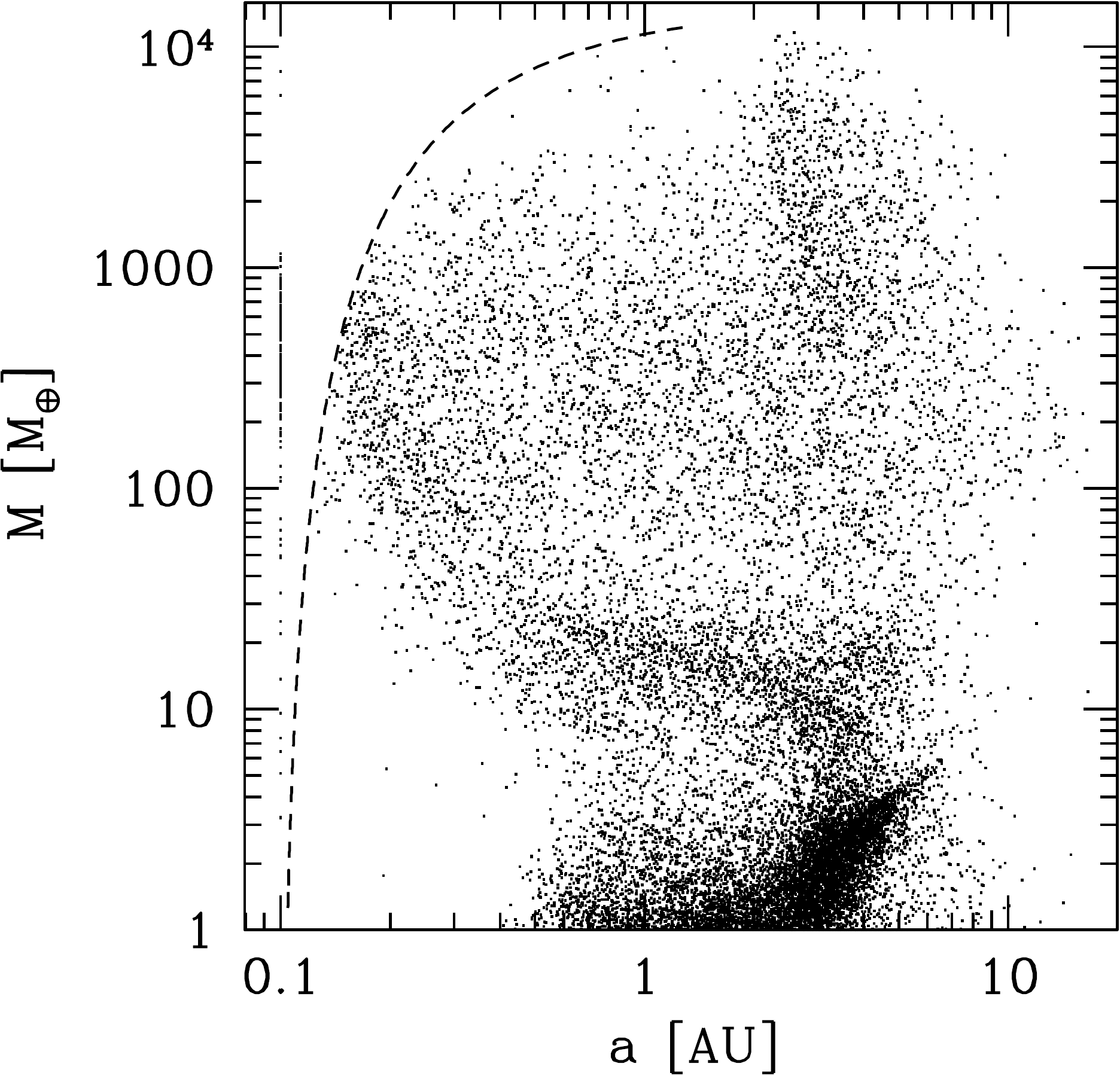}}
      \caption{Planetary population obtained assuming that no planetesimals exist inside the parts of the disk where the initial temperature is larger than 1600 K. The ``rockline effect'' leads to the formation of many Hot Jupiters, but almost completely eliminates Hot Neptunes, which leads to a completely different IMF of the ``Hot'' planets. } 
      \label{fig:amwithrockline}
\end{figure}

In figure \ref{fig:mhotf1dists} we have also plotted the mass histogram of the planets in the feeding limit for another population which is identical to the nominal case (\textit{i.e.} $\f1=0.001$) except that we have included the rockline (Paper I). The corresponding population is plotted in fig. \ref{fig:amwithrockline}. It illustrates how populations synthesis can be used to study the global consequences of a given theoretical description of some physical mechanism, and to reject it if it leads to a population that is in disagreement with the observed planets.

In this simulation it is thus assumed that at distances where the disk midplane temperature is higher than 1600 K at $t=0$ (the moment where the disk evolution starts) no planetesimals exist. This means that inside the rockline $\arock$, the solid surface density drops to zero. This affects both the generation of the initial conditions, as well as the formation of some planets later on. For the initial conditions it means that the minimal distance from where embryos can start must be larger than the rockline.

During the formation of the planets on the other hand, the following interesting mechanism is observed as predicted by Papaloizou \& Terquem (\cite{PT99}): Subcritical planets of the ``horizontal branch'' migrating in planet dominated type II get to the rockline, where they suddenly become supercritical, as the solid accretion rate drops to zero (Pollack et al. \cite{pollacketal1996}; Alibert et al. \cite{alibertetal2005b}). This leads to a rapid accretion of gas, and therefore to a completely different mass histogram of the ``Hot'' planets, with a strong peak at about 1-2 $\mj$, but almost no Hot Neptunes. At the small distance of a few 0.1 AU, the mass of  the rapidly growing planet quickly overcomes the local disk mass (Paper I), so that the migration rate goes down into the increasingly slow, planet dominated mode, which prevents many planets from migrating further in.  An inner hole in the solid surface density therefore leads automatically to a pile up of about Jupiter mass planets at a distance which lies somewhere inside the position of the rockline $\arock$ as seen by the migrating planet. This is very well seen in fig. \ref{fig:amwithrockline} where the ``horizontal branch'' bends upwards at a distance of about 0.4 AU, the location of the rockline for mean disk masses (Paper I). The ``rockline effect'' could therefore be a disk thermodynamics controlled, coupled formation and also stopping mechanism for the observed Hot Jupiter population. Note that the ``rockline effect''   
leads to a mass-distance (anti-)correlation (the mass decreases with increasing distance) qualitatively reminiscent of  the observed correlation pointed out by Mazeh et al. (\cite{mazehetal2005}).

The difficulty is to specify the \textit{relevant} location (for the planetary formation process) of the rockline, as  the disk temperature profile so close to the star changes very rapidly at the beginning of the evolution of the disk. The rockline initially moves quickly inwards, at least in an $\alpha$ disk without irradiation like the one we employ in our model, so that defining the location of the rockline at $t=0$, as it has been done for this test, likely overestimates the importance of  the effect. Observations indeed indicate smaller inner dust disk truncation radii (Eisner et al. \cite{eisneretal2005}), which would shift the ``rockline effect'' closer in. This overestimation is also supported by the fact that the population obtained in this way has so many more close-in Hot Jupiters than Hot Neptunes that it seems incompatible with the recent detections of several Hot Neptune and Super Earth planets (e.g. Udry et al. \cite{udryetal2006}). The HARPS high precision program indicates that there are many more low mass, close-in planets than Hot Jupiters (Mayor et al. \cite{mayoretal2008}), and also in the Keck data many such (not yet announced) candidates exist (Cumming et al. \cite{cummingetal2008}). Indeed, when we again separate the ``Hot'' planets of this special population into two mass bins as in the previous section, we find that now there are 15 times more massive planets than planets with a mass less than 30 $\mearth$ which is inconsistent with these observations. We therefore conclude that modeling the possible absence of solids close to the star in the mentioned way is inappropriate, and must be replaced by a different theoretical description. 

In reality, planetesimals drift inwards due to gas drag and thus might follow the instantaneous location of the rockline, depending on the timescales of planetesimal drift versus rockline recession, an effect that we do not include in the model. The question whether the ``rockline effect'' can occur in nature depends thus on how the timescales of disk temperature evolution, planetesimal drag, and embryo arrival at small distances compare to each other. In any case, the mass histogram of the ``Hot'' planets (as the CoRoT and Kepler mission will provide) can tell us a lot about migration mechanisms, but also about the thermodynamic structure of the disk in proximity to the star.  

\section{Predictions for extremely precise RV surveys}\label{sect:predictionsRV}
In this second part of the result section, we return to the nominal synthetic population and use it to make predictions for the results of radial velocity surveys with an extreme precision. It also serves as an illustration how theoretical populations synthesis calculations can be used to estimate the impact of certain instrumental properties of a given detection technique. 

\subsection{RV measurements at the 1 m/s and the 0.1 level}
Up to this point, we have always used a radial velocity precision equal $\epsilonmc=10$ m/s and a survey length of $\tmc=10$ years as an appropriate mean representation of past real extrasolar planet search programs. In the last few years however, detections mainly made with the HARPS spectrometer (Pepe et al. \cite{pepeetal2004}) made it clear that an RV precision of 1 m/s or even better has become possible, and that  0.1 m/s should be possible too in future even though some issues as the intrinsic oscillations of stars must be overcome.

These improvement have allowed to start exploring new, exciting parts of the planetary IMF in the Neptunian and Super Earth mass domain (e.g. Santos et al. \cite{santosetal2004}; McArthur et al. \cite{mcarthuretal2004}; Lovis et al. \cite{lovisetal2006}; Udry et al. \cite{udryetal2007}; Mayor et al. \cite{mayoretal2008}). Here we study how the properties of the detectable extrasolar population change if $\epsilon_{\rm RV}$ goes down to 1 m/s or even 0.1 m/s.

\subsection{Detectable sub-population}\label{subsect:obspopulationRV}
\begin{figure*}[!ht]
     \centering
      \includegraphics[width=18cm]{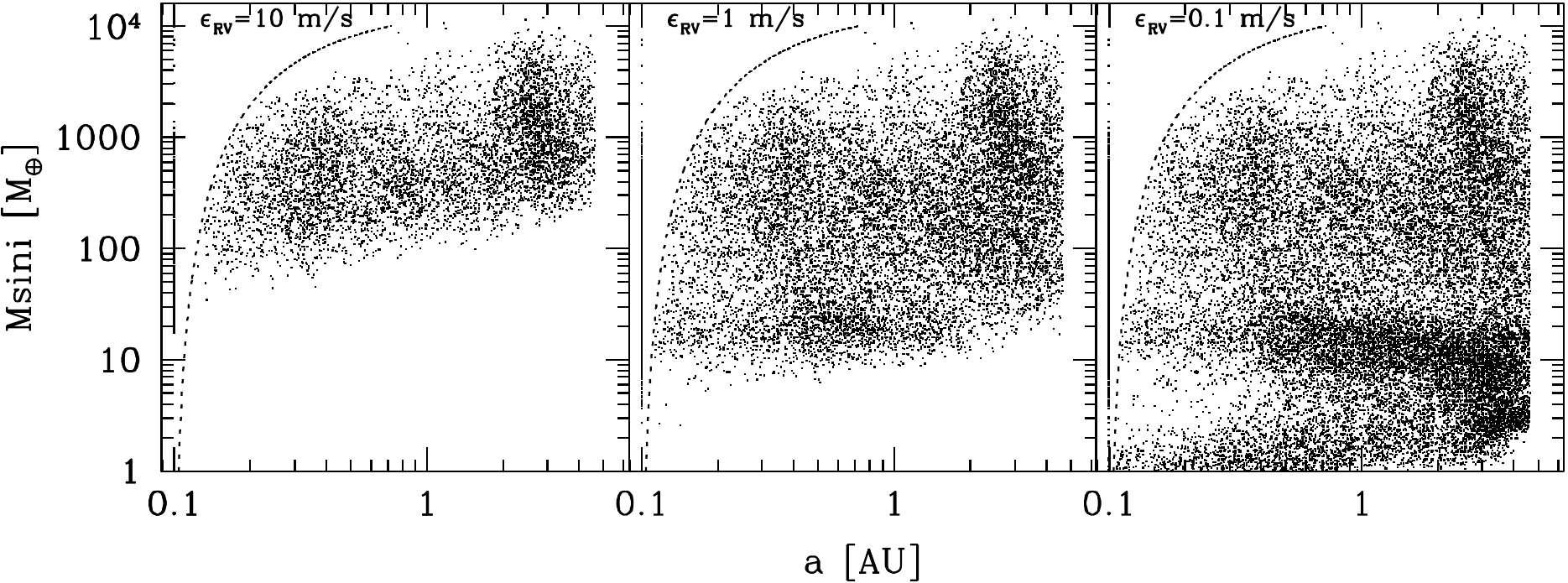}
      \caption{The detectable sub-population for an assumed instrumental precision $\epsilonmc$ of 10, 1 and 0.1 m/s. In all cases, the survey duration $\tmc$ is 10 years. It is likely that the population of Earth mass planets appearing at 0.1 m/s is in reality more abundant than shown in the plot, where the number of such planets at small distances is just given by the fraction of disks with an isolation mass larger than $\membstart$ inside the iceline. Even so, the plot illustrates very well how many extrasolar planet remain to be discovered.} 
      \label{fig:aM3RVacc}
\end{figure*}

In fig. \ref{fig:aM3RVacc} we have plotted the $a-\msini$ diagram for a RV precisions of the synthetic survey of 10 m/s (same as panel B in fig. \ref{fig:amdist}), 1 m/s and 0.1 m/s. The survey duration $\tmc$ is in all cases 10 years. Note that the synthetic detection bias, which uses real data from the ELODIE survey, leads to a reduction of the detection probability already at an induced radial velocity amplitude that is larger than the detection threshold, but allows with a low probability also the detection of planets below it, which is particularly well visible in the 0.1 m/s case, where many low mass planets with an amplitude of the order of 0.1 m/s exist. Fig. \ref{fig:aM3RVacc} makes it obvious that the lower $\epsilonmc$, the closer the detectable population gets to the underlying full population. 

At each precision, new types of planets that become detectable put new constraints on planet formation theories.

At 10 m/s, mainly Jovian planets are detected. Their frequency allows us to extract constraints on the timescales on which sufficiently massive cores must be built. Hot Jupiters and other giant planets at intermediate distances show that planetary migration is an important process which must be included in formation models. The ``metallicity effect'' among these Jovian planets is a strong indication that core accretion is the dominant giant gaseous planet production channel, as this trend is naturally reproduced by many formation models based on this paradigm (\S \ref{subsect:metallicityeffect}). The diversity of giant planets can be interpreted as the consequence of disk properties. Finally, the dependence of the overall maximal mass at a given semimajor axis constrains disk structures and stopping mechanisms, especially if  a very large number of stars (Ge \cite{ge2007}) is observed over several years. 

At 1 m/s, the detection of many Neptunian planets in the upper part of the ``horizontal branch'' out to several AU  becomes possible, as well as the discovery of some Super Earth planets at distances  below 1 AU. It also becomes possible to observationally determine the amount of depletion in the ``planetary desert''. This constrains the timescale of gas runaway accretion, and/or the exact efficiency with which gas accretion continues after a gap has been opened by a planet. The upper boundary of the ``horizontal branch'' indicates the mass where planets go into gas runaway accretion (30-40 $\mearth$).

At 0.1 m/s, almost the complete planetary population, especially also Super Earth and terrestrial planets of $\gtrsim 1$ $\mearth$ at $\lesssim1$ AU will become detectable. The maybe most fundamental implication of the calculations presented here, namely that almost all stars with no (giant) planet detectable today should harbor low mass planets instead (tab. \ref{tab:planettypesbymassfractions}) will be observationally put to the test.  Additionally, at such a precision, the lower part of the population of subcritical Neptunian planets migrating inwards in the ``horizontal branch'' also becomes easily detectable, and will give interesting hints on the criteria for planets to migrate in disk dominated type II migration (planetary mass, disk scale height, gap opening criteria). It also becomes possible to determine if  the planetary IMF is indeed trimodal or if this changed by late time growth. Finally, the abundance of planets with a mass of a few $\mearth$ inside 0.3 AU will help to constrain type I migration rates (\S\ref{subsubsect:f1variation}).
 
\begin{figure}
      \centering
  \resizebox{\hsize}{!}{\includegraphics{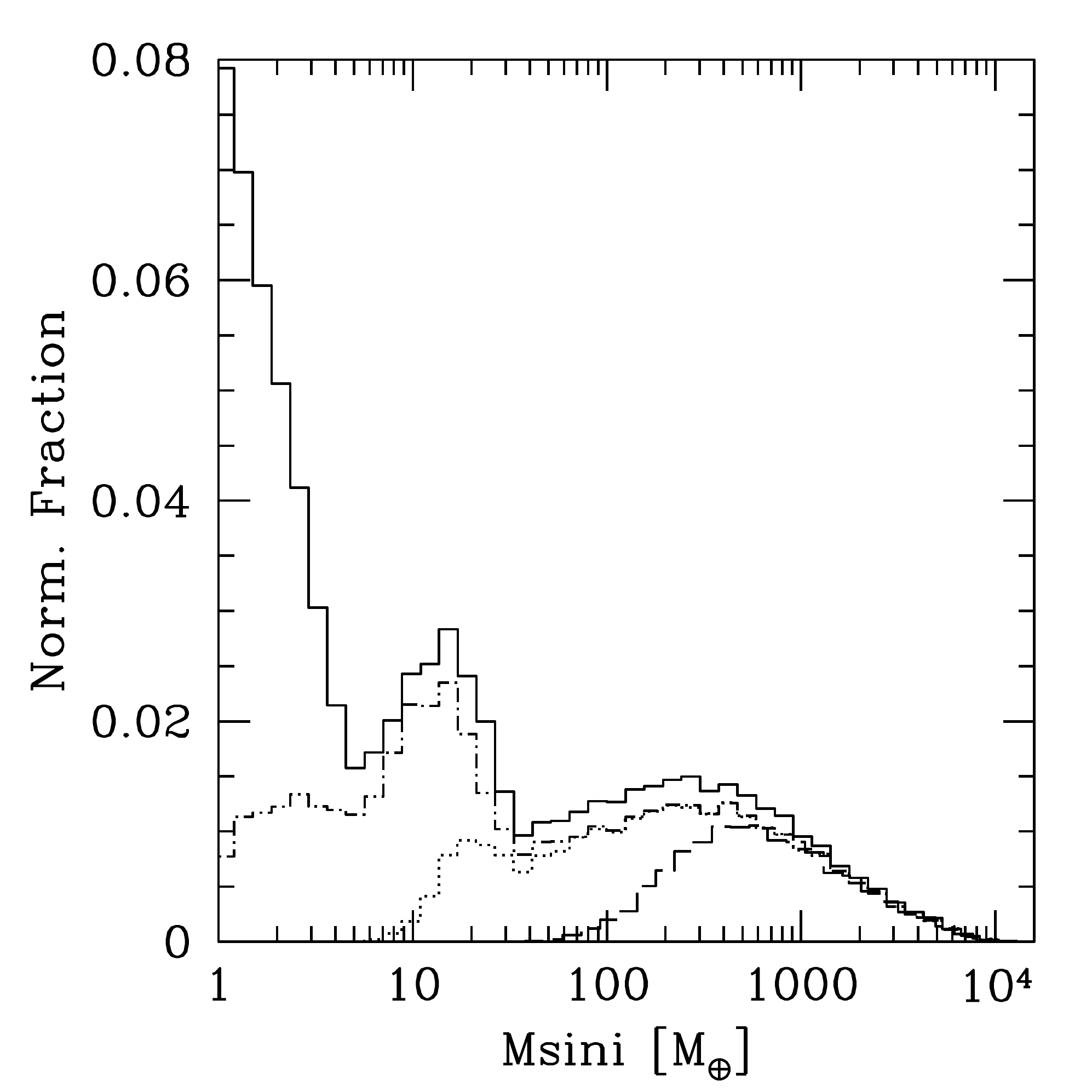}}
      \caption{Distribution of $\msini$. Solid line: Full population. Dashed, dotted, dash-dotted lines: Detectable sub-populations of a 10 year RV survey at 10, 1 and 0.1 m/s, respectively. All curves are normalized by the number of planets  $\nsynt$ in the full population. Note how the distribution becomes bimodal at a precision of 1 m/s.} 
      \label{fig:msinihisto3PRV}
\end{figure}

Figure \ref{fig:msinihisto3PRV} shows the distribution of the projected masses of the full planetary population (corresponding to the IMF in fig. \ref{fig:imf}) together with the mass histogram of the detectable sub-populations in fig. \ref{fig:aM3RVacc} at 10, 1 and 0.1 m/s. The distribution at 10 m/s was compared to the observational comparison sample in fig. \ref{fig:msinidist}. At 10 m/s, the mass distribution has only one maximum, and a relatively simple shape. At 1 m/s in contrast, the minimum in the IMF at 30-40 $\mearth$ and the maximum at $10-20$ $\mearth$ becomes visible, making the distribution bimodal. Interestingly, very recent discoveries at the precision of $\lesssim1$ m/s (Mayor at al. \cite{mayoretal2008}; Mayor \& Udry \cite{mayorudry2008}) also indicate such a shape of the mass function at the lowest masses we can currently detect. The comparison with the results presented here must however be done very carefully, as effects of the primary mass (many very low mass planets orbit around M stars) and of planetary multiplicity (most such very low mass planets are in multiple planetary systems) could be at work, too. 

Compared to the distribution at 10 m/s we see that the 1 m/s distribution carries much more information (and thus constraints) on the planetary formation process, like here on the effect of gas runaway accretion (\S \ref{subsubsect:planetaryIMF}). The 0.1 m/s curve finally must  be interpreted keeping in mind all the caveats concerning the incompleteness of the model at very low masses. This applies for example to the predicted trimodal shape of the distribution. A solid prediction of the model remains however  that many Super Earth planets should be detected: At 0.1 m/s, planets with a mass less than 7 $\mearth$ make up $\sim25\%$ of all detected planets, and $\sim52\%$ have a mass less than 30 $\mearth$.   

Note that in fig. \ref{fig:msinihisto3PRV} the histogram for e.g. 0.1 m/s still lies below the distribution of the full population even at large masses. This is simply due to planets with a period longer than 10 years.

\subsection{Survey detection threshold}\label{subsect:detectionprobability}
\begin{figure}
   \resizebox{\hsize}{!}{\includegraphics{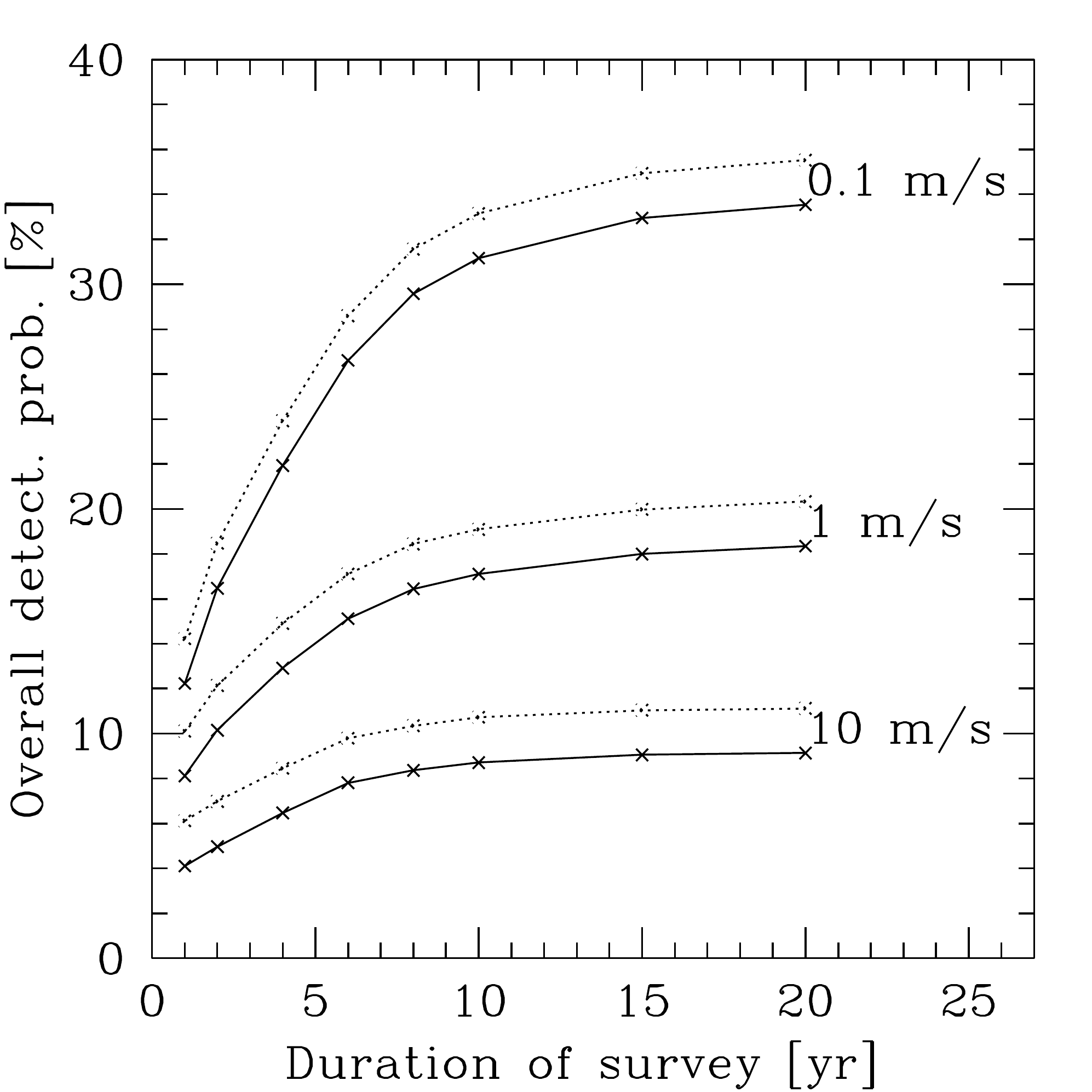}}
   \caption{Overall detection probabilities, \textit{i.e.} the fraction of embryos that become detectable planets, for the synthetic  survey as a function of its time duration $\tmc$ for three different precisions $\epsilonmc$. The solid line assumes that all planets reaching $\atouch$ fall into the star ($P$), while the dotted line assumes that the get detectable, 
regardless of their actual mass at $\atouch$ ($\pwhot$). The curves for 0.1 m/s and to a lesser extent also the one for 1 m/s must be considered as lower limits.} 
   \label{fig:RVyield}
\end{figure}

After varying $\epsilonmc$ only, we can also ask how the overall detection probability of the synthetic survey changes as a function of its duration $\tmc$.  In figure \ref{fig:RVyield},  the overall detection probabilities without  and with ``Hot'' planets ($P$ and $\pwhot$, respectively)  are plotted as function of the synthetic survey's duration, again for $\epsilonmc=$10, 1 and 0.1 m/s.  These fractions can be used as an estimation for the detection yield of real surveys, although that the one-embryo-per-disk limitation (\S\ref{subsect:definingobsconstarints}) is likely to turn the number of detectable planets artificially down. If we can use the Solar System with its four terrestrial planets as a guideline, then the effect should be particularly pronounced at low masses.

In general, as expected, the longer the survey, and the better the instrumental precision, the higher the number of detected planets. First, the detection probability increases quickly with ongoing observations, as the semimajor axis up to which planets are found increases.  After some time ($\approx$10, 20 yr for $\epsilonmc$=10, 0.1 m/s, respectively) only few new detections occur and the curve flattens out. This can be understood by inspection of the $a-\msini$ diagram: For $\epsilonmc=10$ m/s, it is mainly the fact that beyond $\sim 3$ AU, the maximal planet mass starts to decrease again. For  $\epsilonmc=0.1$ m/s, the reason is the paucity of planets outside $\sim7$ AU.

The level at which the detection threshold stalls is $\sim9$ -  11\% for 10 m/s. At 1 m/s precision, the detection rate flattens out at around  19 - 20 \%. For the 0.1 m/s accuracy planets are found around 34 - 36 \% of the observed stars after 20 years of observation, and the saturation is not yet completely reached. This latter values, and to a smaller extent also the 1 m/s predictions, are lower limits only. The fact that we nevertheless find that it should be possible to detect at 0.1 m/s extrasolar planets around at least every third star is promising for future extreme RV precision instruments like ESPRESSO or CODEX (Pepe \& Lovis \cite{pepelovis2008}; Cristiani et al. \cite{cristianietal2007}). Note that the yield of such an extreme precision RV survey should already start at high values ($\sim20\%$ after 3 years).  

It is interesting to compare the synthetic results with some predictions based on observed data and extrapolation to longer periods: Marcy et al. (\cite{marcyetal2005}) have extrapolated the total rate of occurrence of giant planets to be roughly 12\% which is consistent with the number of  long term radial velocity drifts in their data. Naef et al. (\cite{naefetal2005}), by inverting the detection probability map of the ELODIE survey deduce a fraction of stars with planets larger than 0.47 $\mj$ and a period smaller than 3900 d of $7.3 \pm1.5\%$. We find a somewhat larger fraction of synthetic planets fulfilling these criteria of 11.4 \%.  In a similar study, Cumming et al. (\cite{cummingetal2008}), find a occurrence rate of planets with masses $0.3<\msini/\mj<15$ of $\sim14\pm2\%$ inside 10 AU, using a flat extrapolation to larger semimajor axes. In the synthetic population, 9172/70\,000=13.1 \% of the planets fall in this category, in good agreement. Outside 10 AU however, we predict that the occurrence rate does no more raise a lot, as can deduced from the unbiased IMF (tab. \ref{tab:planettypesbymassfractions}): The three highest mass bins ($M>100$ $\mearth$) make up 14.3 \% of all planets.  The reason for this is that in our nominal model ($a-M$ plot in fig. \ref{fig:amdist}), the highest mass planets can reach at $\sim10$ AU is $\mmax\sim10$ $\mj$, but rapidly decreasing to $\sim100$ $\mearth$ at $\sim20$ AU. In Paper I a number of possibilities were discussed that could influence the behavior of $\mmax(a)$. Future observations, possibly with techniques especially apt in detecting massive planets at large distances as astrometry (e.g. Unwin et al. \cite{unwinetal2007}) or direct imaging (e.g. SPHERE, Beuzit et al. \cite{beuzitetal2007}) will provide important insights and help to make our understanding of planet formation more complete.

Developing reliable synthetic observational biases for these two techniques, as well as for transit measurements, which can then be coupled to synthetic planetary populations as demonstrated here for the RV technique will give very important stimulus to better understanding planet formation, as different detection techniques often have the ability to constrain different aspects of the planetary formation process in a complementary fashion. This is the case for the microlensing technique (Cassan \& Kubas \cite{cassankubas2007}) which has already demonstrated its ability to detect very low mass planets (Beaulieu et al. \cite{beaulieuetal2006}) in a part of the mass-orbit plane which is not accessible to current RV surveys. Here we just note that these discoveries fit well into our predictions that many very low mass planets exist at intermediate semimajor axes out to $\sim10$ AU.       

\section{Summary and conclusions}\label{sect:conclusions}
We have used our extended core accretion model to synthesize populations of extrasolar planets orbiting solar type stars. We have identified the subset of potentially detectable synthetic planets and compared them with a sub-sample of 32 actual extrasolar giant planets selected to satisfy the model assumptions. 

The subset of potentially detectable planets has been identified using a RV detection bias model that takes into account the intrinsic instrumental precision as well as the duration of observational surveys. To keep the model tractable, we have adopted a precision of 10 m/s and a survey length of 10 years as representative.

We find that the synthetic survey has an overall detection probability of 8.7 \%, in good agreement with the actual result. The total fraction of initial conditions that eventually lead to the formation of a planet more massive than 100 $\mearth$ is  14.3 \%, again in good agreement with the observationally extrapolated total fraction of stars with a giant planet (Cumming et al. \cite{cummingetal2008}). 

We have then made several Kolmogorov-Smirnov tests to compare the synthetic and observed distributions of some of the most important characteristics of the known planetary population. 

To compare whether observed and synthetic planets of similar masses are located at similar semimajor axes, we use a two dimensional KS test in the mass-orbit plane. We find a significance of 87.8 \% that both the synthetic and the observed planets are drawn from the same parent distribution. In particular, the lack of massive planets at small semimajor axes in the observations (Zucker \& Mazeh \cite{zuckermazeh2002}) is found as well in the models. 

The comparison of the mass distributions alone leads to a high KS significance of 95.6 \%. The distribution has a peak at about 1-2 $\mj$. The decrease at larger masses corresponds to a decrease in underlying, unbiased distribution, while the decrease towards smaller masses is a combined consequence of the decrease of the underlying population and the detection bias. 

We have studied in detail this underlying planetary IMF, which is characterized by several minima and maxima with physical significance. The synthetic IMF has a global maximum at the lowest mass ($\sim1$ $\mearth$) that can occur in in our model, a next local maximum at about 15 $\mearth$, and a third at about 1-2 $\mj$. The first minimum at $\sim7$ $\mearth$ corresponds to the transition between Super Earth and Neptunian planets. Note that since our PIMF corresponds to the mass distribution at the time the gaseous disk vanishes, it could be possible to fill-up this minimum by the merging of smaller mass planets.  The next minimum at 30 $\mearth$ is more robust and separates solid dominated planets from giant gaseous planets that undergo runaway gas accretion (Ida \& Lin \cite{idalin2004a}). It is very characteristic of the core accretion mechanism. At large masses ($\gtrsim1000$ $\mearth$) the PIMF decrease with increasing mass. However, a long tail up to almost 40 $\mj$ exists.

The comparison of the semimajor axis leads to a KS significance of 63.9 \%. We also find an raise of the distribution after a initially flat part in $\log(a)$, but in the model it occurs at about 2 AU, rather than about 1 AU as in the observational data (Udry \& Santos \cite{udrysantos2007}). We interpret the fact that we reproduce the mass distribution better than the semimajor axis as an indication that we describe mass accretion in a less rudimentary way than migration. It is also probable that the semimajor axis is more affected than the mass distribution by the consequences of N-body interactions after disk dispersal which are not included in the model (Thommes et al. \cite{thommesetal2008}).

At last, we have compared the metallicity distributions. We find that, similarly to observations (Santos et al. \cite{santosetal2004}), the detectable synthetic planets are characterized by a metallicity distribution that is shifted towards larger [Fe/H] by about 0.1 dex compared to the initial distribution. However, a rather low significance of about 22 \% is returned by the KS test. We suspect that a reason for this low significance could be connected to the extremely simple planetesimal disk model used in these calculations. Despite this,  the ``metallicity effect'', which is the observed increase of the detection probability with stellar metallicity is also present in the synthetic population. Even if the increase of the detection probability in the model seems to follow a slightly different (weaker) dependence on [Fe/H], we satisfy the observational constraint within its error bars (Fischer \& Valenti \cite{fischervalenti2005}). 

The next observational constrain we studied is the fraction of stars with a Hot Jupiter in orbit. Planets inside 0.1 AU have been excluded from the previous quantitative comparisons, as the knowledge of the disk properties and processes close to the star are very uncertain. Assuming now for simplicity that the mass of the planets remains constant once they have reached the inner limit of our computational disk ($\sim0.1$ AU), we find a rate of occurrence of Hot Jupiters of about 0.4-0.6 \%, compatible with observations (Fressin et al. \cite{fressinetal2007}). For these ``Hot'' synthetic planets, there is a positive correlation between the stellar metallicity and the maximal total amount of heavy elements they contain, in agreement with internal structure modeling for transiting exoplanets (Guillot et al. \cite{guillot2008}). We have thus satisfied constraints coming from transit surveys, too. Our population synthesis also shows that the formation timescales of giant planets (mean formation timescale of 3.5 Myr) are in agreement with observed disk lifetimes which is another important constraint coming from the formation environment itself (Haisch et al. \cite{haischetal2001}). 

Our conclusions from these comparisons is that the core accretion paradigm coupled with migration, even implemented in a much simplified manner, can generate a planet population with characteristics resembling the one of the actually detected ones. The crucial point for this conclusion is that the synthetic population can reproduce a large number of different observational constraints \textit{simultaneously}. 

This was not evident from the beginning as a large number of combinations of model parameters did not lead to acceptable results. For example,  the efficiency of type I migration needed to be reduced considerably (10 to a 1000 times)  from the linear rate by Tanaka et al. (\cite{Tanaka}) otherwise it would have been impossible to reproduce the observed semimajor axis distribution (but see also Ida \& Lin \cite{idalin2008b}). An other example concerns the possible absence of solids inside the rockline which would be an efficient mechanism to produce close-in giant planets, which could explain the observed pile-up.  We believe that such results and constraints are the essence of the utility of the population synthesis efforts presented here.  

Finally, the nominal population allows us to make predictions about the planets which currently cannot be detected. As an illustrative example, we have shown the impact of improving the precision at which radial velocities can be measured. Our results indicate that the observed mass distribution becomes bimodal at a RV precision of 1 m/s. The latest discoveries indeed point to such a feature (Mayor \& Udry \cite{mayorudry2008}). At a precision of even 0.1 m/s, a large fraction of the underlying planetary population will become detectable (planets will be found around at least 30-40 \% of all FGK stars), and the observed mass function traces the characteristics of the underlying actual mass function. At a time when new high precision instruments are in the planing, this could be used to quantitatively define the instrument requirements need to achieve a specific science goal. 

\acknowledgements
We thank Stephane Udry, Isabelle Baraffe, Gilles Chabrier, Nuno Santos and Tsevi Mazeh for useful discussions. We are thankful for interesting and valuable comments by an anonymous referee. This work was supported in part by the Swiss National Science Foundation. Computations were made on the ISIS, ISIS2 and UBELIX clusters at the University of Bern, and on the cluster of the Observatoire de Besan\c{c}on funded by the Conseil G\'en\'eral de Franche-Comt\'e.

\end{document}